\documentclass{article}

\PassOptionsToPackage{numbers, compress}{natbib}
\usepackage[main, final]{neurips_2026}
\usepackage[utf8]{inputenc}
\usepackage[T1]{fontenc}
\usepackage{hyperref}
\usepackage{url}
\usepackage{booktabs}
\usepackage{amsfonts}
\usepackage{nicefrac}
\usepackage{microtype}
\usepackage{xcolor}

\usepackage{amsmath}
\usepackage{mathtools}
\usepackage{amsthm}
\usepackage{algorithmic}

\usepackage{algorithm}
\usepackage{graphicx}
\usepackage{subcaption}
\usepackage{xspace}
\usepackage{amssymb}

\usepackage{wrapfig}

\theoremstyle{plain}
\newtheorem{theorem}{Theorem}[section]
\newtheorem{proposition}[theorem]{Proposition}
\newtheorem{lemma}[theorem]{Lemma}

\theoremstyle{definition}
\newtheorem{definition}[theorem]{Definition}

\theoremstyle{remark}

\newcommand{\ours}{MCGI\xspace}

\title{MCGI: Manifold-Consistent Graph Indexing for Billion-Scale Disk-Resident Vector Search} 
 
\author{
  Dongfang Zhao \\
  University of Washington, Tacoma School of Engineering and Technology \\
  \texttt{dzhao@uw.edu}
}

\begin{document}

\maketitle 

\begin{abstract}
Graph-based Approximate Nearest Neighbor (ANN) search often suffers from performance degradation in high-dimensional spaces due to the Euclidean-Geodesic mismatch, where greedy routing diverges from the underlying data manifold. To address this challenge, this paper presents Manifold-Consistent Graph Indexing (MCGI), a geometry-aware and disk-resident indexing method that leverages Local Intrinsic Dimensionality (LID) to allocate the search effort of a graph index according to the intrinsic geometry of data. Specifically, MCGI estimates the LID around each query from the distances that its own search has already computed and resizes the candidate list of the query once, so that queries in high-LID regions receive longer lists while queries in low-LID regions terminate earlier; the estimate itself costs no additional disk reads. Theoretical analysis shows that the success probability of any fixed search list decays exponentially with the LID of the query. At matched high recall, \ours{} raises the query throughput of DiskANN on the same standard Vamana index by up to $1.43\times$ on three million-scale benchmarks and by up to $1.24\times$ on the billion-scale SIFT1B and Deep1B.
\end{abstract}

\section{Introduction}
\label{sec:intro}

The advent of Large Language Models~\cite{brown2020language, touvron2023llama} has fundamentally transformed the landscape of information retrieval and knowledge management. To address the inherent limitations of LLMs, such as hallucinations~\cite{ji2023survey} and knowledge cutoff dates, Retrieval-Augmented Generation (RAG)~\cite{lewis2020retrieval} has emerged as a critical architectural paradigm. RAG relies heavily on the ability to retrieve semantically relevant context from massive corpora in real-time, typically via dense vector representations~\cite{karpukhin2020dense}. This dependency has placed Approximate Nearest Neighbor Search (ANNS) at the core of modern data infrastructure, demanding vector indices that can scale to billion-point datasets~\cite{simhadri2022neurips} while maintaining low latency and high recall under strict production constraints.

For extreme-scale embedding vector applications, many ANNS techniques have noticeably converged on graph-based indices, with DiskANN (Vamana) and its variants~\cite{subramanya2019diskann,osdi25pipeann,fast26odinann} being representative solutions for large-scale, SSD-resident workloads. These algorithms typically employ greedy routing on a proximity graph to navigate from an entry point to the query target. While such methods exhibit exceptional performance on popular benchmarks like SIFT1M~\cite{jegou2011product} (128 dimensions), their efficiency degrades significantly on data of high intrinsic dimensionality. For example, GloVe-100 and GIST1M have mean estimated local intrinsic dimensionalities of 35.3 and 48.9 against 16.1 for SIFT1M (Table~\ref{tab:search_params}), and DiskANN serves three to four times fewer queries per second on them than on SIFT1M at 90\% recall (Figure~\ref{fig:results}). This degradation is often attributed to the curse of dimensionality~\cite{bellman1957dynamic}, where the distance contrast diminishes, and the Euclidean shortest path on the graph diverges from the geodesic path on the underlying data manifold~\cite{tenenbaum2000global,NEURIPS2023_d0ac28b7}. We refer to this phenomenon as the \emph{Euclidean-Geodesic Mismatch}, or EGM. When the routing algorithm ignores the intrinsic geometry of the data, it performs excessive backtracking and disk I/O, rendering the search inefficient, which is particularly problematic in high-recall scenarios where production systems often expect Recall@10 higher than 95\%.

Our observation on this EGM problem is that high-dimensional real-world data is rarely uniformly distributed; Instead, it typically adheres to the Manifold Hypothesis~\cite{tenenbaum2000global,roweis2000nonlinear}, residing on lower-dimensional structures embedded within the ambient space. Consequently, the search difficulty is not uniform across the dataset but depends on the Local Intrinsic Dimensionality (LID)~\cite{NIPS2004_74934548}. That is, in regions where the data manifold is flat (low LID), greedy routing is effective; however, in regions with high curvature or complex topology (high LID), standard greedy strategies may fail to identify the descent direction. 
\emph{We thus argue that an optimal indexing strategy must be manifold-aware, dynamically allocating computational resources based on the local geometric complexity.}

To demonstrate the above hypothesis, we design Manifold-Consistent Graph Indexing (MCGI), a geometry-aware disk-based indexing method that allocates the resources of a graph index according to the LID of the data. During search, MCGI estimates the LID around each query from the distances that its search has computed after a few tens of expansions, and it resizes the candidate list of the query once, so that queries in high-LID regions receive longer lists and the other queries stop earlier. This search-time adaptation changes neither the index format nor the I/O path of the search; consequently, it applies to any Vamana index and removes disk reads that a fixed list would spend on easy queries. The same rule can also set the pruning parameter of each node during construction, where a higher LID relaxes the pruning test of the node. 

\emph{The remainder of this paper will present the following contributions:} (1) We establish a theoretical framework that relates the local intrinsic dimensionality to graph navigability, with a formal guarantee on fixed-beam suboptimality; (2) We design geometry-aware algorithms that apply one LID-driven allocation rule at both stages of a disk-resident graph index, namely a query-adaptive candidate list that sizes the search of each query by the LID around it and, as an optional construction-time component, a pruning parameter that relaxes the pruning test of high-LID nodes; and (3) We evaluate MCGI on five benchmarks and show up to $1.33\times$, $1.43\times$, and $1.20\times$ higher throughput than DiskANN at matched high recall on SIFT1M, GloVe-100, and GIST1M, respectively, up to $1.22\times$ and $1.24\times$ on one billion vectors of SIFT1B and Deep1B, respectively, and 13--18\% fewer disk reads per query than DiskANN and PipeANN on GIST1M.

\section{Related Work}
\label{sec:related_work}

\paragraph{Vector Indexing Paradigms}
While traditional sparse retrieval methods rely on lexical matching~\cite{robertson2009probabilistic}, memory-resident graph indices like HNSW~\cite{malkov2018efficient} establish state-of-the-art dense retrieval but face billion-scale memory bottlenecks. Disk-based approaches, such as DiskANN~\cite{subramanya2019diskann}, SPANN~\cite{chen2021spann}, PipeANN~\cite{osdi25pipeann}, and OdinANN~\cite{fast26odinann}, along with predecessors like NSG~\cite{fu2019nsg}, adapt topologies or inverted structures for SSDs. Recent advancements such as CSPG~\cite{NEURIPS2024_bab1486c} further optimize sparse proximity graphs  to reduce unpromising explorations, yet these methods predominantly rely on static routing parameters evaluated on moderate dimensionality. In contrast, MCGI adjusts the search of each query to the local topology to align graph traversal with the underlying manifold.

\paragraph{High-Dimensional Indexing}
Early space-partitioning trees like the KD-tree~\cite{bentley1975multidimensional} and R-tree~\cite{guttman1984r} severely degrade to linear-scan performance in dimensions higher than 20 due to the curse of dimensionality. Approximate methods address this through sub-linear hashing like LSH~\cite{indyk1998approximate, datar2004locality}, which incurs excessive storage overhead, or subspace quantization like PQ~\cite{jegou2011product} and OPQ~\cite{ge2013optimized}, and randomized structures like RP-Trees~\cite{dasgupta2008random}, which sacrifice accuracy through compression. Furthermore, recent literature extensively studies the theoretical limits of navigable graphs~\cite{NEURIPS2024_6dc63b40} and the efficiency of index construction~\cite{10.14778/3594512.3594527} in strictly high-dimensional scenarios. MCGI keeps the graph traversal of a disk-resident index, which uses compressed codes only to order candidates and full-precision vectors from disk for the final distances, and adapts the length of this traversal to the local topology of each query.

\paragraph{Intrinsic Dimensionality}
Theoretical metrics such as the doubling dimension~\cite{gupta2003bounded} and expansion dimension~\cite{karger2002finding} provide asymptotic bounds on search complexity, and recent analyses establish worst-case performance limits for popular implementations like DiskANN and HNSW based on intrinsic dimensionality~\cite{NEURIPS2023_d0ac28b7}. Maximum likelihood estimators~\cite{NIPS2004_74934548} for Local Intrinsic Dimensionality (LID) enable robust practical estimations. Existing works typically utilize LID passively for query hardness prediction~\cite{he2012difficulty, 10.14778/3704965.3704974} or detecting adversarial examples~\cite{ma2018characterizing} without altering the underlying index structure. Diverging from passive descriptive applications, MCGI uses the LID of each node and of each query as a control quantity, which sets the pruning parameter during construction and the candidate-list size during search.

\section{Methodology} 
\label{method}

Greedy routing needs more steps to find a descent direction where the local intrinsic dimensionality (LID) around the target is high. \ours{} therefore spends more of the resources of a disk-resident graph index where the LID is high and fewer where it is low. During search, \ours{} gives a query with a high LID a longer candidate list and a query with a low LID a shorter one, because a fixed list fails on the first kind of query and wastes disk reads on the second. During construction, \ours{} can apply the same rule to the pruning parameter of each node: a node with a high LID receives a larger pruning parameter, which relaxes the pruning test for its candidate edges. Under the degree bound of the index, nodes typically fill their neighbor lists with either parameter; this construction-time rule thus mainly changes the choice of edges of a node. The search-time rule applies to any Vamana index, regardless of how the index was built.

\subsection{Mapping Function from LID to Pruning Parameter}
\label{subsec:mapping}

We start by briefly reviewing the notions of Local Intrinsic Dimensionality (LID) in the words of analysis. A full treatment can be found, for example, in Houle~\cite{DBLP:conf/sisap/Houle17}.

\begin{definition}[Local Intrinsic Dimensionality]
\label{def:lid}
    Let $\mathcal{X}$ be a domain equipped with a distance measure $d: \mathcal{X} \times \mathcal{X} \to \mathbb{R}^+$. For a reference point $x \in \mathcal{X}$, let $F_x(r) = \mathbb{P}(d(x, Y) \le r)$ denote the cumulative distribution function (CDF) of the distance between $x$ and a random variable $Y$ drawn from the underlying data distribution, where $\mathbb{P}$ denotes the probability function and $r$ denotes the radius. 
    The Local Intrinsic Dimensionality (LID) of $x$, denoted as $\text{LID}(x)$, is defined as the intrinsic growth rate of the probability measure within the neighborhood of $x$:
    \begin{equation}\label{eq:pid}
        \text{LID}(x) \triangleq \lim_{r \to 0} \frac{r \cdot F'_x(r)}{F_x(r)} = \lim_{r \to 0} \frac{d \ln F_x(r)}{d \ln r},
    \end{equation}
    provided that the limit exists and $F_x(r)$ is continuously differentiable for $r > 0$.
\end{definition}

While Definition~\ref{def:lid} is elegant, it is often impractical to compute directly. According to~\cite{10.1145/2783258.2783405}, LID can be estimated as follows.

\begin{definition}[LID Maximum Likelihood Estimator]
\label{def:lid_mle}
Consider a reference point $x$ and its $k$ nearest neighbors under the distance measure $d$, and denote by $r_i = d(x, v_i)$ the distance to the $i$-th nearest neighbor, sorted such that $r_1 \le \dots \le r_k$. 
Following the formulation in~\cite{10.1145/2783258.2783405}, which adapts the Hill estimator for intrinsic dimensionality, the LID at $x$ can be estimated as:
\begin{equation}
    \widehat{\text{LID}}(x) = - \left( \frac{1}{k} \sum_{i=1}^{k} \ln \frac{r_i}{r_k} \right)^{-1}.
\end{equation}
\end{definition}

The primary hypothesis of the proposed Manifold-Consistent Graph Indexing (MCGI) is that the graph topology should adapt to the local geometric complexity, rather than being a static vertex-edge structure. 
To preserve topological fidelity in complex regions, the indexing algorithm must adopt a conservative pruning strategy there (larger $\alpha$ in~\cite{subramanya2019diskann}), which relaxes the pruning test for the candidate edges of each node.
The question then becomes how to quantitatively map the estimated LID to the pruning parameter $\alpha$ that controls the edge selection during graph construction.
We refer to such a function as the \textit{mapping function} $\Phi: \mathbb{R}^+ \to [\alpha_{\min}, \alpha_{\max}]$, which translates the geometric complexity of the local neighborhood into a corresponding pruning constraint.

Formally, consider a node $u \in V$, where $V$ denotes the vertex set of the graph, i.e., the indexed data points, and denote by $\widehat{\text{LID}}(u)$ its estimated LID. We define the \emph{pruning parameter} $\alpha(u)$ as:
\begin{equation}
    \alpha(u) \triangleq \Phi( \widehat{\text{LID}}(u) ),
\end{equation}
where $\Phi$ will be constructed to satisfy the following geometric requirements: in regions with high LID, the graph should prune conservatively (larger $\alpha$) to retain the navigational edges that greedy routing needs; conversely, in low-LID regions, where greedy routing finds descent directions easily, the pruning can be aggressive (smaller $\alpha$) and keep the graph sparse.

To ensure the mapping is robust across datasets with varying complexity scales, we employ Z-score normalization based on the empirical distribution of the LID estimates. We first compute the normalized score $z(u)$:
\begin{equation}
    z(u) = \frac{\widehat{\text{LID}}(u) - \mu_{\widehat{\text{LID}}}}{\sigma_{\widehat{\text{LID}}}},
\end{equation}
where $\mu_{\widehat{\text{LID}}}$ and $\sigma_{\widehat{\text{LID}}}$ denote the mean and standard deviation of the set of estimated LID values $\{ \widehat{\text{LID}}(v) \mid v \in V \}$ computed across the entire graph.
We then formulate $\Phi$ using a logistic function to smoothly map the Z-score to the operational range $[\alpha_{\min}, \alpha_{\max}]$:
\begin{equation}
    \Phi(\widehat{\text{LID}}(u)) = \alpha_{\min} + \frac{\alpha_{\max} - \alpha_{\min}}{1 + \exp(-z(u))}.
\end{equation}


The mapping function $\Phi$ satisfies the following geometric properties that are essential for stable graph construction: Monotonicity and Boundedness. We state these properties in the following as propositions with proof sketches and provide full proofs in the Appendix.

\begin{proposition}[Monotonicity]
\label{prop:monotonicity}
The mapping function $\Phi$ is strictly increasing with respect to the estimated local intrinsic dimensionality. Formally, given that the standard deviation of the LID estimates $\sigma_{\widehat{\text{LID}}} > 0$ and the pruning range $\alpha_{\max} > \alpha_{\min}$, the derivative satisfies:
\begin{equation}
    \frac{d \Phi}{d \widehat{\text{LID}}(u)} > 0.
\end{equation}
\end{proposition}

\begin{proof}[Proof Sketch]
By the chain rule, $\frac{d\Phi}{d\widehat{\text{LID}}(u)} = \frac{d\Phi}{dz} \cdot \frac{dz}{d\widehat{\text{LID}}(u)}$, where the Z-score satisfies $\frac{dz}{d\widehat{\text{LID}}(u)} = \frac{1}{\sigma_{\widehat{\text{LID}}}} > 0$.
The derivative of the logistic component evaluates to $\frac{d\Phi}{dz} = (\alpha_{\max} - \alpha_{\min}) \cdot \frac{e^{-z}}{(1+e^{-z})^2}$, which is strictly positive since $\alpha_{\max} > \alpha_{\min}$ and $e^{-z} > 0$ for all $z \in \mathbb{R}$.
Combining these two terms yields a strictly positive derivative: $\alpha(u)$ increases monotonically with the geometric complexity.
\end{proof}

\begin{proposition}[Boundedness]
The pruning parameter $\alpha(u)$ derived from the mapping function is strictly bounded within the prescribed operational interval. That is, for any node $u$ with a finite LID estimate, the following holds:
\begin{equation}
    \alpha_{\min} < \alpha(u) < \alpha_{\max}.
\end{equation}
\end{proposition}

\begin{proof}[Proof Sketch]
For any finite $\widehat{\text{LID}}(u)$, the Z-score $z(u)$ is finite; consequently, the logistic function satisfies $S(u) = \frac{1}{1+e^{-z(u)}} \in (0, 1)$ strictly, since $e^{-z(u)} \in (0, \infty)$.
Substituting into $\alpha(u) = \alpha_{\min} + (\alpha_{\max} - \alpha_{\min}) \cdot S(u)$ and applying the strict bounds on $S(u)$ directly yields $\alpha_{\min} < \alpha(u) < \alpha_{\max}$.
\end{proof}

\subsection{Indexing Algorithms}
\label{subsec:algorithm}


The Manifold-Consistent Graph Indexing (MCGI) algorithm introduces a geometric calibration phase to a graph indexing procedure. Unlike static methods that apply a uniform connectivity rule, MCGI executes in two distinct stages to ensure that the topology respects the manifold structure, as follows. The full pseudo-code of the algorithm is in Appendix~\ref{appendix:alg}, Algorithm~\ref{alg:mcgi}.

\paragraph{Phase 1: Geometric Calibration.}
Before modifying the graph topology, the system first performs a global analysis of the dataset geometry. We estimate the LID for every point and aggregate the population statistics ($\mu, \sigma$) defined in Section~\ref{subsec:mapping} once before the construction starts, so that the mapping function $\Phi$ stays fixed and cheap to evaluate during the edge-selection process.

\paragraph{Phase 2: Manifold-Consistent Refinement.}
The index construction then refines the neighbor set $\mathcal{N}(u)$ of each node $u$ in the graph $G$ over several iterations, where each iteration updates $\mathcal{N}(u)$ from a candidate pool $\mathcal{C}$ that a greedy search on the current graph returns:
\begin{enumerate}
    \item Queries the pre-computed geometric profile to determine the node-specific constraint $\alpha(u)$.
    \item Explores the graph to identify a candidate pool $\mathcal{C}$.
    \item Filters connections using the dynamic pruning criterion. 
\end{enumerate}

\paragraph{Remark} The calibration phase estimates the LID of each point from its approximate $k$ nearest neighbors in $O(N \log N)$ time, which the construction phase dominates, on par with the complexity of DiskANN~\cite{subramanya2019diskann}.
In practice, calculating the exact LID for the entire dataset may cause a computational bottleneck on billion-scale data. 
To address this, we also design an online algorithm called \textit{Online-MCGI}, whose pseudo-code can be found in Appendix~\ref{appendix:alg}, Algorithm~\ref{alg:online_mcgi}.



\subsection{Theoretical Analysis}
\label{sec:theory}

\subsubsection{Geometric Complexity and Adaptive Routing}
\label{subsec:geometric_complexity}

The efficiency of greedy routing on a proximity graph is intrinsically tied to the local topological properties of the underlying manifold $\mathcal{M}$. Under the Manifold Hypothesis, the search space is locally characterized by the Local Intrinsic Dimensionality (LID) at query $q$. 
We formalize the relationship between LID and the computational cost required to identify the nearest neighbor, as follows.

\begin{lemma}[Local Complexity Lower Bound]
\label{lemma:complexity}
Consistent with the complexity bounds established for growth-restricted metrics~\cite{karger2002finding}, consider a query $q$ on a manifold $\mathcal{M}$ of local intrinsic dimensionality $d = \text{LID}(q)$ under the assumptions that the neighbor directions at a node are independent and uniform on the unit sphere, that the step length is proportional to the distance to $q$, and that each greedy step must reduce this distance by a fixed factor $1-\epsilon$ with $\epsilon$ independent of $d$. Then the expected number of distance evaluations $N_{dist}$ per improving step scales exponentially with $d$:
\begin{equation}
    \mathbb{E}[N_{dist}] \ge \Omega\left( \sqrt{d} \cdot \exp(\lambda \cdot d) \right),
\end{equation}
where $\lambda > 0$ depends on $\epsilon$ and the ratio of the step length to the distance.
\end{lemma}

\begin{proof}[Proof Sketch]
At a node $u$ with distance $r$ to query $q$, a greedy step succeeds only if a neighbor $v$ falls within a cone of half-angle $\theta_{\max}$ around the direction of $q$; the probability of this event corresponds to the ratio of a spherical cap to the full hypersphere $\mathcal{S}^{d-1}$.
Bounding the numerator of this ratio from above by integration by parts and its denominator from below near the equator yields $P_{\text{success}} \le \kappa_0\, d^{-1/2} \exp(-\lambda \cdot d)$ for a constant $\kappa_0$, where the exponential decay arises from the concentration of measure on high-dimensional spheres and $\lambda = -\ln(\sin\theta_{\max}) > 0$.
Since $\mathbb{E}[N_{dist}] = 1/P_{\text{success}}$, the lower bound follows directly.
\end{proof}

Lemma~\ref{lemma:complexity} reveals a fundamental geometric barrier: the probability of a successful greedy step decays exponentially with the LID, so that a fixed budget $L$ fails more often in high-LID regions. To address this, \ours{} adopts an \textit{Iso-Recall} strategy, aiming to homogenize the search reliability across the heterogeneous manifold.

\begin{proposition}[Optimal Budget Allocation]
\label{thm:oba}
To maintain a uniform bound on the routing failure probability $\delta$ across the manifold (i.e., $\mathbb{P}(\text{fail}\;|\;q) \le \delta$ for all $q \in \mathcal{M}$), the search beam width $L(q)$ must grow exponentially with the local intrinsic dimensionality:
\begin{equation}
    L(q) = \Omega\left( \exp\left( \lambda \cdot \text{LID}(q) \right) \right).
\end{equation}
\end{proposition}

\begin{proof}[Proof Sketch]
For a beam of width $L$, the condition $(1 - P_{\text{success}})^L \leq \delta$ holds exactly when $L \geq \ln(1/\delta) / (-\ln(1-P_{\text{success}}))$, which approaches $\ln(1/\delta)/P_{\text{success}}$ for small $P_{\text{success}}$.
Because one expansion evaluates at most $R$ neighbors, Lemma~\ref{lemma:complexity} and the union bound give $P_{\text{success}} = O(R\, d^{-1/2}\exp(-\lambda \cdot d))$, and $\sqrt{d} \ge 1$ yields $L(q) = \Omega(\exp(\lambda \cdot \text{LID}(q))/R)$.
\end{proof}

Proposition~\ref{thm:oba} implies that the search budget of each query should grow with the LID of its neighborhood. A change of $L$ at every expansion step would disrupt memory alignment and the vectorized distance computation of the search loop; in contrast, a single resize of the candidate list per query leaves both intact. \ours{} therefore realizes the budget of Proposition~\ref{thm:oba} at query time through the query-adaptive search list of Section~\ref{subsec:query_adaptive}; in addition, the pruning parameter $\alpha(u)$ can complement this budget at construction time.


\subsubsection{Topological Fidelity and Connectivity}
\label{subsec:connectivity}

A primary theoretical concern with aggressive edge pruning is the potential fracture of the connectivity backbone~\cite{dzhao_pami09}. We show that, when the pruning test runs over all points and no degree bound applies, the pruning of \ours{} preserves every edge of the Euclidean minimum spanning tree and therefore global reachability.

Formally, the edge selection in \ours{} is governed by the adaptive pruning parameter $\alpha(u)$. Specifically, with the candidates of $u$ processed in increasing order of their distance to $u$, an edge $(u, v)$ is pruned if a previously selected neighbor $n$ of $u$ satisfies $\alpha(u) \cdot d(n, v) \le d(u, v)$.
Because such a witness also satisfies $d(u, n) < d(u, v)$, this condition defines an \textit{exclusion region} $\{n : d(u,n) < d(u,v),\ d(n,v) \le d(u,v)/\alpha(u)\}$. As $\alpha(u)$ increases, this region shrinks, making pruning more conservative. We formulate this in the following proposition.

\begin{proposition}[Connectivity Preservation]
\label{prop:connectivity}
Let $G_{EMST}$ and $G_{RNG}$ denote the Euclidean Minimum Spanning Tree (EMST) and the Relative Neighborhood Graph (RNG), respectively. For any configuration of points with distinct pairwise distances, provided that $\alpha(u) \ge 1.0$ for all $u \in V$, that the pruning test of each node runs over all other points, and that no degree bound truncates the neighbor lists, the graph $G_{MCGI}$ satisfies the inclusion hierarchy:
\begin{equation}
    E_{EMST} \subseteq E_{RNG} \subseteq E_{MCGI}.
\end{equation}
Consequently, $G_{MCGI}$ is connected.
\end{proposition}

\begin{proof}[Proof Sketch]
The first inclusion $E_{EMST} \subseteq E_{RNG}$ follows from the classical result of Toussaint~\cite{TOUSSAINT1980261}, which holds for any finite point set in a metric space.
For the second inclusion $E_{RNG} \subseteq E_{MCGI}$, note that MCGI prunes an edge $(u,v)$ only with a witness $n$ that satisfies $d(u,n) < d(u,v)$ and $d(n,v) \le d(u,v)/\alpha(u) \le d(u,v)$ since $\alpha(u) \geq 1.0$; such a witness lies in the RNG exclusion region, so any edge surviving the RNG pruning condition survives MCGI pruning at both of its endpoints.
Connectivity of $G_{MCGI}$ follows immediately from the inclusion hierarchy, since $G_{EMST}$ spans all vertices by definition.
\end{proof}

Proposition~\ref{prop:connectivity} ensures that there are no structural dead ends. While Lemma~\ref{lemma:complexity} dictates the \textit{cost} of routing, Proposition~\ref{prop:connectivity} ensures the \textit{viability} of routing. Even in the worst-case scenario where heuristics fail, a path exists from the entry point to any target via the edges of the underlying EMST, provided the graph traversal algorithm has sufficient width to discover these backbone links.

\subsubsection{Fixed-Beam Suboptimality and Adaptation}
\label{subsec:optimality}

Proposition~\ref{thm:oba} shows that a common recall target requires a beam width that grows exponentially with the LID of the query. Queries with different LIDs therefore need different list sizes. The following definition makes this need precise by assigning each query the list size that keeps its failure probability within a common bound; all queries then reach the same recall. A fixed list size is then too short for the queries that need a longer list and wastes disk reads on the queries that need a shorter one. Proposition~\ref{prop:fixed_beam} bounds the success probability of a query under a fixed list size by the ratio of this size to the list size of the definition.

\begin{definition}[Iso-Recall Beam Schedule]
\label{def:iso_recall}
Consider a query $q$ and a beam search toward $q$, and model each expansion of the search as an independent trial that finds a descent step with probability $P_{\mathrm{success}}(q) \in (0, 1)$. A candidate list of size $L$ then fails to find a descent step with probability $(1 - P_{\mathrm{success}}(q))^{L}$, which we require to be at most a target $\delta \in (0, 1)$. Since $1 - x \le e^{-x}$ for every real $x$, the failure probability satisfies $(1 - P_{\mathrm{success}}(q))^{L} \le \exp(-L \cdot P_{\mathrm{success}}(q))$. The requirement therefore holds whenever $\exp(-L \cdot P_{\mathrm{success}}(q)) \le \delta$. Taking the logarithm of both sides, this condition reads $-L \cdot P_{\mathrm{success}}(q) \le \ln \delta$, i.e., $L \ge \ln(1/\delta)/P_{\mathrm{success}}(q)$. The \emph{iso-recall beam schedule} assigns the query $q$ the smallest list size that satisfies this condition:
\begin{equation}
    L^*(q) = \frac{\ln(1/\delta)}{P_{\mathrm{success}}(q)}.
\end{equation}
A search algorithm is \emph{recall-optimal} if it uses the list size $L^*(q)$ for every query $q$.
\end{definition}

The list size $L^*(q)$ grows exponentially with the LID of $q$, because the success probability $P_{\mathrm{success}}(q)$ in its denominator, which counts one expansion rather than one neighbor, decays exponentially with the LID. Specifically, one expansion evaluates at most $R$ neighbors, where $R$ is the maximum out-degree of the index. Lemma~\ref{lemma:complexity} bounds the probability that one neighbor yields a descent step by $\kappa_0\, \mathrm{LID}(q)^{-1/2} \exp(-\lambda \cdot \mathrm{LID}(q))$, where $\lambda > 0$ and $\kappa_0 > 0$ are the constants of the lemma. The union bound over these at most $R$ neighbors, together with $\mathrm{LID}(q) \ge 3$, then gives
\begin{equation}
    L^*(q) \ge C_\delta \cdot \exp\!\left(\lambda \cdot \mathrm{LID}(q)\right), \qquad C_\delta = \frac{\ln(1/\delta)}{R\, \kappa_0}.
    \label{eq:oracle_schedule}
\end{equation}

\begin{proposition}[Suboptimality of Fixed Beam]
\label{prop:fixed_beam}
For any fixed beam width $\bar{L}$, the probability that the search succeeds at a query $q$ is at most $\ln(1/\delta)\, \bar{L}/L^*(q)$, where $L^*(q)$ is the iso-recall list size of Definition~\ref{def:iso_recall}. Since $L^*(q)$ grows exponentially with the LID of $q$, this probability decays exponentially with the LID.
\end{proposition}

\begin{proof}[Proof Sketch]
Since the search succeeds only if at least one of its $\bar{L}$ expansions succeeds, the union bound gives a success probability of at most $\bar{L}\, P_{\mathrm{success}}(q)$. Substituting $P_{\mathrm{success}}(q) = \ln(1/\delta)/L^*(q)$ from Definition~\ref{def:iso_recall} yields the bound, whose exponential decay follows from Eq.~\eqref{eq:oracle_schedule}.
\end{proof}

Proposition~\ref{prop:fixed_beam} explains why static-parameter
methods degrade on high-dimensional data: the success probability of
a fixed beam decays exponentially with the LID of the query. A list
long enough for a typical query therefore fails on the queries in
high-LID regions. The query-adaptive search list of Section~\ref{subsec:query_adaptive} reduces this gap directly by approximating $L^*(q)$ for each query.

\subsection{Query-Adaptive Search List}
\label{subsec:query_adaptive}

\ours{} applies the iso-recall schedule of Definition~\ref{def:iso_recall} at query time by resizing the candidate list of each query according to the local geometry around the query. The schedule requires $\mathrm{LID}(q)$ before the search starts, although the neighborhood of $q$ is unknown at that point. The beam search of a disk-resident Vamana index, however, keeps two structures that reveal this neighborhood during the search: the candidate list holds up to $\bar{L}$ nodes ordered by their compressed distances to $q$, whereas the expanded set holds the exact distances of all nodes read from disk so far. \ours{} therefore inspects these two structures once after $n_0$ expansions, with $n_0$ a small constant of a few tens, and resizes the candidate list at this checkpoint.

At the checkpoint, \ours{} forms a feature vector $\mathbf{f}(q)$ of five quantities that describe the local geometry around $q$. The first feature is $\ln \widehat{\mathrm{LID}}(q)$ from Definition~\ref{def:lid_mle} over the exact distances $r_1 \le \dots \le r_k$ of the $k$ nearest expanded nodes, with $k$ equal to the result size, typically 10. Because $k$ distances yield a noisy estimate, the second feature applies the same estimator to the compressed distances of the $m$ best candidates in the list, with $m$ a few tens. The next two features record the scale of the neighborhood through $\ln r_1$ and $\ln r_k$, since the LID estimator removes this scale by construction. The last feature is the logarithm of one plus the number of candidates whose squared distance lies within $(1+\epsilon)\, r_k^2$ for a small constant $\epsilon$ such as 0.2. A large count means that many points compete for the last result positions, as happens when a high LID concentrates the distances.

\ours{} maps the features to a list size through a log-linear score that mirrors the exponential form of Eq.~\eqref{eq:oracle_schedule}. The score of a query $q$ is $\mathbf{w}^{\top}\mathbf{f}(q) + b$, whose weights $\mathbf{w}$ and offset $b$ result from a least-squares fit with a small ridge penalty on a set of calibration queries disjoint from the evaluation queries; the regression target of each calibration query is the logarithm of the smallest list size at which the query reaches its best recall. The constant $\bar{s}$ denotes the mean score of the calibration queries. Given the list size $\bar{L}$ that the application requests, \ours{} scales $\bar{L}$ by the exponential of the centered score times a strength $\gamma$ and clips the result to the range from the result size $k$ to $c\bar{L}$:
\begin{equation}
    L(q) = \min\Bigl\{ c\bar{L},\; \max\bigl\{ k,\; \bar{L} \exp\bigl(\gamma\,(\mathbf{w}^{\top}\mathbf{f}(q) + b - \bar{s})\bigr) \bigr\} \Bigr\}.
    \label{eq:adaptive_list}
\end{equation}
Table~\ref{tab:adaptive_params} lists the meaning, the source, and the typical value of every quantity in Eq.~\eqref{eq:adaptive_list} and in the features; only the strength $\gamma$ is chosen, on the calibration queries.

\begin{table}[t]
\centering
\small
\caption{Quantities of the query-adaptive search list.}
\label{tab:adaptive_params}
\begin{tabular}{llll}
\toprule
Symbol & Meaning & Source & Typical value \\
\midrule
$\bar{L}$ & List size that the application requests & Application & -- \\
$k$ & Result size; lower bound of the list size & Application & 10 \\
$n_0$ & Expansions before the checkpoint & Fixed & 24 \\
$m$ & Candidates in the candidate-list LID estimate & Fixed & 64 \\
$\epsilon$ & Margin of the competition count & Fixed & 0.2 \\
$\mathbf{f}(q)$ & Five geometric features at the checkpoint & Search & -- \\
$\mathbf{w}, b$ & Weights and offset of the score & Calibration fit & -- \\
$\bar{s}$ & Mean score of the calibration queries & Calibration queries & -- \\
$\gamma$ & Strength; $\gamma = 0$ gives the fixed list size $\bar{L}$ & Calibration choice & 0.5--1.3 \\
$c$ & Cap factor; bounds the disk reads of one query & Fixed & 8 \\
\bottomrule
\end{tabular}
\end{table}
A query of average geometry has a centered score of zero and therefore keeps the requested list size $\bar{L}$. A query whose features indicate a high LID receives a longer list, whereas a query in a low-LID region terminates earlier and saves the disk reads that a fixed list would spend on it.

The adaptation adds no disk reads, because every feature reuses distances that the search has already computed. Computing $\mathbf{f}(q)$ takes $O(n_0 + \bar{L})$ time once per query, which is small compared with the $O(\bar{L} R)$ compressed-distance evaluations of the search itself. Resizing the list either truncates its tail or raises its capacity, after which the beam search continues from its current state without a restart. The vectorized distance computation therefore remains unchanged, since the adaptation only moves the termination point of each query. Because the index format is also unchanged, the query-adaptive search list applies to any Vamana index, including an index built with the construction-time adaptation of Section~\ref{subsec:algorithm}.

\section{Evaluation}
\label{sec:experiments}

All experiments run on a cloud virtual machine with 192 virtual cores of Intel Xeon Platinum 8468 and 960 GiB of RAM, where the disk-resident indexes reside on a 17.6 TB virtual block device (virtio). Each search reads graph nodes from this device with direct I/O, with a measured I/O time of about 0.34 ms per read at 32 search threads; consequently, the throughput of DiskANN and \ours{}, which share one I/O engine, follows their number of disk reads per query. We evaluate \ours{} on three million-scale benchmarks with dimensionality from 100 to 960, namely SIFT1M~\cite{jegou2011product}, GloVe-100~\cite{pennington2014glove}, and GIST1M~\cite{jegou2011product}, and on the billion-scale SIFT1B~\cite{jegou2011searching} and Deep1B~\cite{babenko2016efficient}. The baselines are the fixed-list search of DiskANN (Vamana)~\cite{subramanya2019diskann} on the same index, the pipelined disk-resident search of PipeANN~\cite{osdi25pipeann}, and the in-memory indexes HNSW~\cite{malkov2018efficient}, IVF-Flat~\cite{johnson2019billion}, and CSPG~\cite{NEURIPS2024_bab1486c}; the latter four run on GIST1M. More setup details are in Appendix~\ref{appendix:setup}.

\subsection{Million-scale Effectiveness}

On each million-scale dataset, \ours{} and the fixed-list search of DiskANN run on one shared index, so that the two differ only in the candidate list of each query. Both methods search an index built with the standard Vamana pruning rule, using the same compressed vectors and the same requested list sizes $\bar{L}$ from 10 to 400. The weights of Eq.~\eqref{eq:adaptive_list} come from 10{,}000 calibration queries per dataset, which are disjoint from the official evaluation queries. Each method runs once per list size on an otherwise idle machine, with the page cache dropped before each run.

\begin{figure*}[t]
    \centering
    \includegraphics[width=\textwidth]{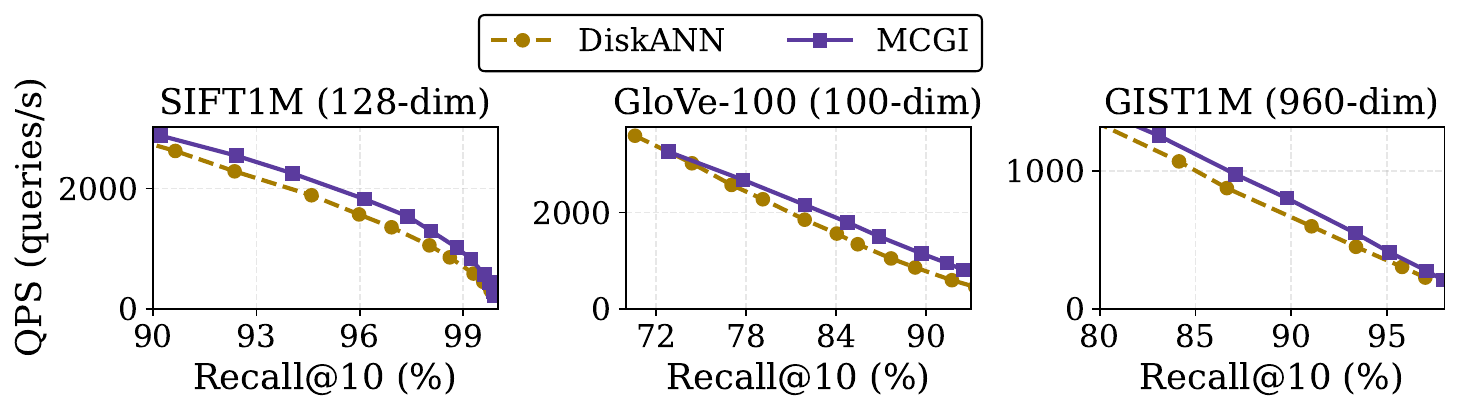}
    \caption{Recall-QPS trade-off of \ours{} and DiskANN on the million-scale datasets.}
    \label{fig:results}
\end{figure*}

Figure~\ref{fig:results} plots the query throughput against Recall@10 for the two methods. \ours{} raises the throughput of DiskANN on all three million-scale datasets, with an improvement that grows with the target recall. On SIFT1M, the gain rises from $1.15\times$ at 95\% recall to $1.33\times$ at 99\%. GloVe-100 and GIST1M follow the same trend, with gains from $1.12\times$ at 80\% to $1.43\times$ at 90\% and from $1.12\times$ at 85\% to $1.20\times$ at 95\%, respectively. The growth with recall agrees with Proposition~\ref{prop:fixed_beam}, because a high target recall forces a fixed search to use a long list for every query, including the many queries in low-LID regions that converge early. \ours{} shortens the lists of these queries and lengthens the lists of queries in high-LID regions, which removes disk reads that do not improve the result.

\subsection{Billion-scale Scalability}

We build one DiskANN index for each of SIFT1B and Deep1B with the standard sharded construction under a memory budget of 300 GiB. The construction splits the 8-bit vectors of SIFT1B into three overlapping shards and completes in 4.8 hours on the CPU cores of the platform, while the 32-bit vectors of Deep1B need seven shards and 5.4 hours. Both methods again search the same index, which uses $R = 32$, $L_{build} = 50$, and 31-byte PQ codes. The calibration queries of each dataset are the even-indexed half of its 10{,}000 official queries, whereas the evaluation uses the odd-indexed half. The requested list sizes extend to $\bar{L} = 1{,}000$ on SIFT1B and to $\bar{L} = 2{,}000$ on Deep1B, since DiskANN needs a list longer than 400 on SIFT1B and longer than 800 on Deep1B to reach 95\% recall.

\begin{figure}[t]
    \centering
    \includegraphics[width=0.67\textwidth]{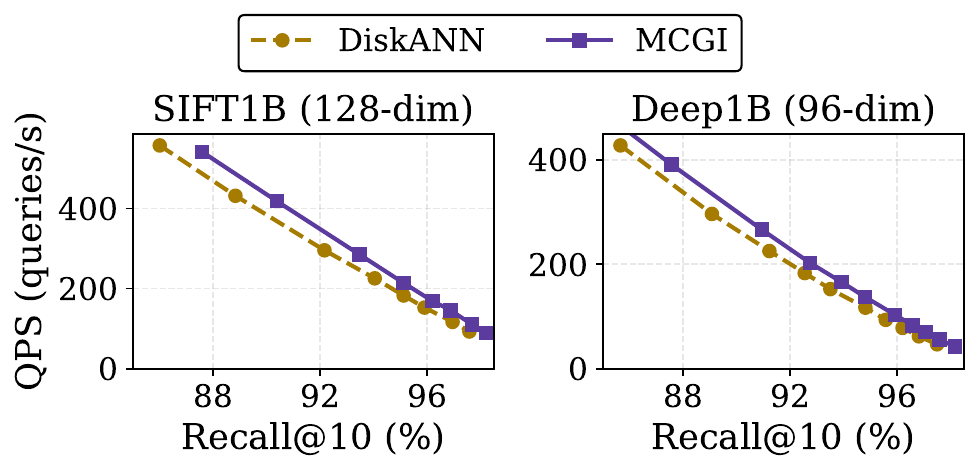}
    \caption{Recall-QPS trade-off of \ours{} and DiskANN on SIFT1B and Deep1B.}
    \label{fig:results_1b}
\end{figure}

Figure~\ref{fig:results_1b} shows that the gain of \ours{} persists at one billion vectors on both datasets and again grows with the target recall. On SIFT1B, \ours{} raises the throughput of DiskANN from $1.07\times$ at 85\% recall to $1.17\times$ at 95\% and to $1.22\times$ at 97\%, where the throughput rises from 115 to 141 queries per second. Deep1B follows the same trend with gains from $1.08\times$ at 85\% to $1.18\times$ at 95\% and to $1.24\times$ at 97\%, where the throughput rises from 58 to 72 queries per second. At the longest list that we evaluate, DiskANN stays below 98\% recall on both datasets, whereas \ours{} reaches 98.2\% on each of them, at 89 queries per second on SIFT1B and 42 on Deep1B.

\subsection{Comparison with Pipelined and In-memory Search}

On GIST1M, two groups of baselines complement the fixed-list search of DiskANN, and both are compared at 95\% Recall@10. The first group is the pipelined disk-resident search of PipeANN~\cite{osdi25pipeann, fast26odinann}, which overlaps the disk reads of each query; PipeANN searches the same DiskANN index files as \ours{} with 32 threads, an in-memory entry-point index built from a 1\% sample, and the better of the I/O widths 4 and 32. The second group consists of the in-memory indexes HNSW~\cite{malkov2018efficient}, IVF-Flat~\cite{johnson2019billion}, and CSPG~\cite{NEURIPS2024_bab1486c}, which keep every full-precision vector in RAM and therefore show the speed that a disk-resident index gives up for its small memory footprint. Table~\ref{tab:gist_compare} covers the first group, whereas Table~\ref{tab:memory_compare} covers the second.

\begin{table}[t]
\centering
\small
\begin{minipage}[t]{0.54\textwidth}
\centering
\caption{QPS and disk reads per query at 95\% recall.}
\label{tab:gist_compare}
\begin{tabular}{lcccc}
\toprule
 & \multicolumn{2}{c}{$R = 32$} & \multicolumn{2}{c}{$R = 48$} \\
\cmidrule(lr){2-3} \cmidrule(lr){4-5}
Method & QPS & Reads & QPS & Reads \\
\midrule
DiskANN & 351 & 272.7 & 659 & 146.7 \\
PipeANN & \textbf{545} & 268.3 & 780 & 143.8 \\
\ours{} & 420 & \textbf{223.7} & \textbf{785} & \textbf{125.5} \\
\bottomrule
\end{tabular}
\end{minipage}
\hfill
\begin{minipage}[t]{0.44\textwidth}
\centering
\caption{QPS and vector RAM at 95\% recall.}
\label{tab:memory_compare}
\setlength{\tabcolsep}{4pt}
\begin{tabular}{lcc}
\toprule
Method & QPS & Vectors in RAM \\
\midrule
IVF-Flat & 1{,}292 & 3.84~GB \\
CSPG ($R = 32$) & 5{,}742 & 3.84~GB \\
HNSW ($M = 16$) & 6{,}678 & 3.84~GB \\
CSPG ($R = 48$) & 9{,}107 & 3.84~GB \\
\ours{} ($R = 48$) & 785 & 0.32~GB \\
\bottomrule
\end{tabular}
\end{minipage}
\end{table}

Against pipelined search, \ours{} issues the fewest disk reads per query, whereas PipeANN gains its speed by overlapping the reads that it issues. Table~\ref{tab:gist_compare} shows that \ours{} reads 13--18\% fewer nodes per query than DiskANN and PipeANN, whose reads differ by at most 2\%. By overlapping its reads, PipeANN reaches a higher throughput with $R = 32$, while \ours{} matches it with $R = 48$. Because the two techniques act on different costs, the query-adaptive list can also size the candidate list of a pipelined search.

The in-memory indexes exceed every disk-resident method in throughput, at the price of a memory footprint that grows with the full-precision vectors. Table~\ref{tab:memory_compare} shows that they serve 1{,}292 to 9{,}107 queries per second, whereas \ours{} serves 785. In exchange, the in-memory indexes hold the 3.84~GB of full-precision vectors in RAM, whereas \ours{} holds mainly the 0.32~GB of PQ codes; this trade-off motivates disk-resident indexes at billion scale.

\subsection{Ablation and Sensitivity}

\begin{figure*}[t]
    \centering
    \begin{minipage}[t]{0.46\textwidth}
        \centering
        \includegraphics[width=\linewidth]{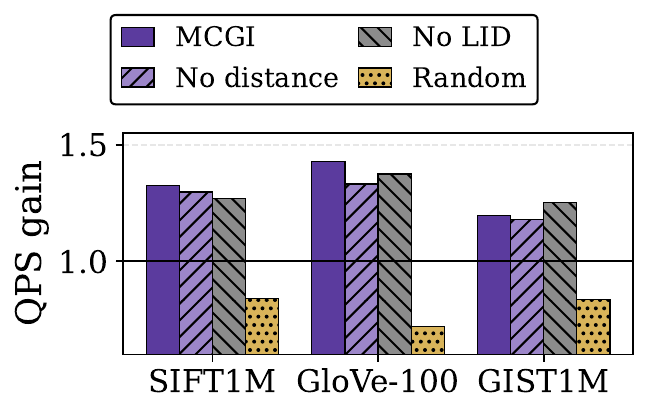}
        \caption{Removing LID or distance features.}
        \label{fig:ablation}
    \end{minipage}
    \hfill
    \begin{minipage}[t]{0.52\textwidth}
        \centering
        \includegraphics[width=\linewidth]{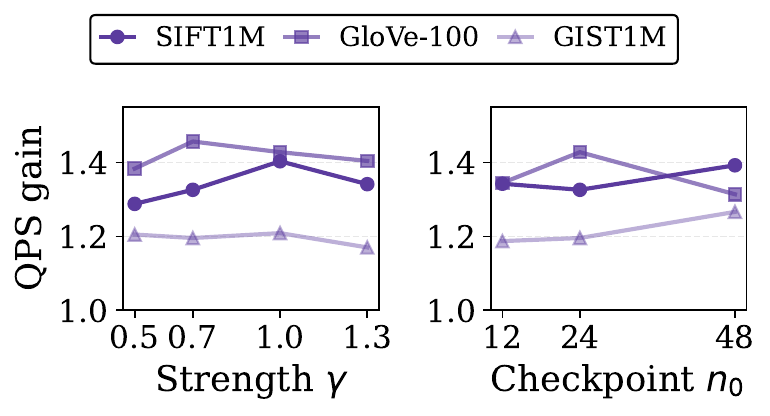}
        \caption{Sensitivity to $\gamma$ and $n_0$.}
        \label{fig:sensitivity}
    \end{minipage}
\end{figure*}

Figure~\ref{fig:ablation} attributes the gain of \ours{} at the highest target recall of each million-scale dataset, namely 99\% on SIFT1M, 90\% on GloVe-100, and 95\% on GIST1M. Each variant is fitted on the same calibration queries and searched once on the same index as DiskANN. The random variant keeps the fitted list sizes of \ours{} in distribution but assigns them to queries at random, which lowers the throughput to $0.72\times$--$0.84\times$ that of DiskANN. The gain therefore comes from matching each list to the geometry of its query, not from varying the list size. Removing a single feature group, in contrast, changes the gain of \ours{} by only a few percent. Without the three distance features, which are the first and tenth expanded distances and the candidate count, the gain falls by 1--7\%, with the largest drop on GloVe-100 from $1.43\times$ to $1.33\times$. Without the two LID features, it falls by 4\% on SIFT1M and GloVe-100 and rises by 5\% on GIST1M. The two groups therefore carry largely the same information, because the LID estimate is a function of these distances.

Figure~\ref{fig:sensitivity} varies the strength $\gamma$ of Eq.~\eqref{eq:adaptive_list} from 0.5 to 1.3 and the checkpoint $n_0$ from 12 to 48 expansions around the setting of Figure~\ref{fig:results}. Every setting keeps a gain of at least $1.17\times$, and the setting of Figure~\ref{fig:results} lies within $0.08\times$ of the best setting on each dataset. Neither parameter therefore needs careful tuning, and the default $n_0 = 24$ serves all datasets.

\section{Conclusion}

This paper presents MCGI, a geometry-aware disk-resident indexing method for high-dimensional data of varying intrinsic dimensionality.
MCGI addresses the Euclidean-Geodesic mismatch by adapting the candidate list of each query, and optionally the pruning parameter of each node, to local intrinsic dimensionality. Our theoretical analysis shows that the success probability of any fixed beam width decays exponentially with the LID of the query, which motivates a query-adaptive list that approximates the iso-recall schedule from distances that the search has already computed. At matched high recall, MCGI raises the throughput of DiskANN by up to $1.43\times$ on three million-scale benchmarks and by up to $1.24\times$ at one billion vectors.


\begin{ack}
Results presented in this paper were obtained using the Chameleon testbed supported by the National Science Foundation.
Copilot and Claude assisted with the system implementation, the experiments, and the writing of this paper; the author verified all results and takes full responsibility for the content. 

The source code of MCGI is available at \url{https://github.com/hpdic/mcgi}. 
\end{ack}

\bibliography{ref}
\bibliographystyle{plain}

\clearpage
\section*{NeurIPS Paper Checklist}

\begin{enumerate}

\item {\bf Claims}
    \item[] Question: Do the main claims made in the abstract and introduction accurately reflect the paper's contributions and scope?
    \item[] Answer: \answerYes{} 
    \item[] Justification: The paper presents both theoretical and experimental evidence to support its claims.

\item {\bf Limitations}
    \item[] Question: Does the paper discuss the limitations of the work performed by the authors?
    \item[] Answer: \answerYes{} 
    \item[] Justification: Appendix~\ref{appendix:limitations} discusses the scope of the evaluated mechanism, the effect of the degree bound on the construction-time rule, the tail latency of the adaptive list, the calibration requirements, and the modeling assumptions of the theory.

\item {\bf Theory assumptions and proofs}
    \item[] Question: For each theoretical result, does the paper provide the full set of assumptions and a complete (and correct) proof?
    \item[] Answer: \answerYes{} 
    \item[] Justification: Every proposition states its assumptions, including the absence of a degree bound where it matters, and Appendix~\ref{appendix:proofs} gives the full proofs; Appendix~\ref{appendix:limitations} lists the modeling assumptions of the analysis.

 \item {\bf Experimental result reproducibility}
    \item[] Question: Does the paper fully disclose all the information needed to reproduce the main experimental results of the paper to the extent that it affects the main claims and/or conclusions of the paper (regardless of whether the code and data are provided or not)?
    \item[] Answer: \answerYes{} 
    \item[] Justification: The code, the experiment scripts, and the data behind every figure are available in the public repository \url{https://github.com/hpdic/mcgi}, and Section~\ref{sec:experiments} and Appendix~\ref{appendix:setup} describe the setup.

\item {\bf Open access to data and code}
    \item[] Question: Does the paper provide open access to the data and code, with sufficient instructions to faithfully reproduce the main experimental results, as described in supplemental material?
    \item[] Answer: \answerYes{} 
    \item[] Justification: The public repository \url{https://github.com/hpdic/mcgi} contains the code, the experiment scripts, and the exported results; all datasets are public benchmarks.

\item {\bf Experimental setting/details}
    \item[] Question: Does the paper specify all the training and test details (e.g., data splits, hyperparameters, how they were chosen, type of optimizer) necessary to understand the results?
    \item[] Answer: \answerYes{} 
    \item[] Justification: Section~\ref{sec:experiments} and Appendix~\ref{appendix:setup} give the datasets, the index and search parameters, the calibration protocol, and the baseline configurations.

\item {\bf Experiment statistical significance}
    \item[] Question: Does the paper report error bars suitably and correctly defined or other appropriate information about the statistical significance of the experiments?
    \item[] Answer: \answerNo{} 
    \item[] Justification: Each configuration runs once on an otherwise idle machine with the page cache dropped, and both methods search the same index with the same queries; the paper does not report error bars.

\item {\bf Experiments compute resources}
    \item[] Question: For each experiment, does the paper provide sufficient information on the computer resources (type of compute workers, memory, time of execution) needed to reproduce the experiments?
    \item[] Answer: \answerYes{} 
    \item[] Justification: Section~\ref{sec:experiments} and Appendix~\ref{appendix:setup} describe the platform (192 virtual cores, 960 GiB of RAM, and a virtual block device), and Table~\ref{tab:offline_cost} reports the build and calibration times.
    
\item {\bf Code of ethics}
    \item[] Question: Does the research conducted in the paper conform, in every respect, with the NeurIPS Code of Ethics \url{https://neurips.cc/public/EthicsGuidelines}?
    \item[] Answer: \answerYes{} 
    \item[] Justification: We believe we have followed the NeurIPS Code of Ethics in all aspects of our research.

\item {\bf Broader impacts}
    \item[] Question: Does the paper discuss both potential positive societal impacts and negative societal impacts of the work performed?
    \item[] Answer: \answerNA{} 
    \item[] Justification: The paper studies the efficiency of nearest-neighbor search over public benchmarks and has no direct societal impact beyond faster retrieval.
    
\item {\bf Safeguards}
    \item[] Question: Does the paper describe safeguards that have been put in place for responsible release of data or models that have a high risk for misuse (e.g., pre-trained language models, image generators, or scraped datasets)?
    \item[] Answer: \answerNA{} 
    \item[] Justification: The paper releases no data or model with a risk of misuse.

\item {\bf Licenses for existing assets}
    \item[] Question: Are the creators or original owners of assets (e.g., code, data, models), used in the paper, properly credited and are the license and terms of use explicitly mentioned and properly respected?
    \item[] Answer: \answerYes{} 
    \item[] Justification: The paper cites the creators of every dataset (SIFT1M, GIST1M, SIFT1B, GloVe-100, and Deep1B) and every baseline. DiskANN and Faiss are released under the MIT license, PipeANN and hnswlib under the Apache License 2.0, and CSPG is used from the public repository of its authors for research comparison; the datasets are public benchmarks used under the terms of their original releases.

\item {\bf New assets}
    \item[] Question: Are new assets introduced in the paper well documented and is the documentation provided alongside the assets?
    \item[] Answer: \answerYes{} 
    \item[] Justification: The code of \ours{} is released at \url{https://github.com/hpdic/mcgi} under the MIT license of its DiskANN code base, with a README that maps every figure and table to its script.

\item {\bf Crowdsourcing and research with human subjects}
    \item[] Question: For crowdsourcing experiments and research with human subjects, does the paper include the full text of instructions given to participants and screenshots, if applicable, as well as details about compensation (if any)? 
    \item[] Answer: \answerNA{} 
    \item[] Justification: The paper involves no crowdsourcing or human subjects.

\item {\bf Institutional review board (IRB) approvals or equivalent for research with human subjects}
    \item[] Question: Does the paper describe potential risks incurred by study participants, whether such risks were disclosed to the subjects, and whether Institutional Review Board (IRB) approvals (or an equivalent approval/review based on the requirements of your country or institution) were obtained?
    \item[] Answer: \answerNA{} 
    \item[] Justification: The paper involves no human subjects.

\item {\bf Declaration of LLM usage}
    \item[] Question: Does the paper describe the usage of LLMs if it is an important, original, or non-standard component of the core methods in this research? Note that if the LLM is used only for writing, editing, or formatting purposes and does \emph{not} impact the core methodology, scientific rigor, or originality of the research, declaration is not required.
    \item[] Answer: \answerNA{} 
    \item[] Justification: LLMs assisted with the implementation, the experiments, and the writing, as stated in the acknowledgments, but they are not a component of the method.

\end{enumerate}

\clearpage
\appendix

\section{Technical appendices and supplementary material}

\subsection{Complete Proofs}
\label{appendix:proofs}

\begin{proposition}[Monotonicity]\label{proof:mono}
The mapping function $\Phi$ is strictly increasing with respect to the estimated local intrinsic dimensionality. Formally, given that the standard deviation of the LID estimates $\sigma_{\widehat{\text{LID}}} > 0$ and the pruning range $\alpha_{\max} > \alpha_{\min}$, the derivative satisfies:
\begin{equation}
    \frac{d \Phi}{d \widehat{\text{LID}}(u)} > 0.
\end{equation}
\end{proposition}

\begin{proof}
Let $L = \widehat{\text{LID}}(u)$ be the independent variable. We define the normalized Z-score $z$ as a function of $L$:
\begin{equation}
    z(L) = \frac{L - \mu_{\widehat{\text{LID}}}}{\sigma_{\widehat{\text{LID}}}}.
\end{equation}
The mapping function is defined as:
\begin{equation}
    \Phi(L) = \alpha_{\min} + \frac{C}{1 + \exp(-z(L))},
\end{equation}
where $C = \alpha_{\max} - \alpha_{\min}$. Since we strictly require $\alpha_{\max} > \alpha_{\min}$, it follows that $C > 0$.
To determine the sign of the gradient, we apply the chain rule:
\begin{equation}
    \frac{d \Phi}{d L} = \frac{d \Phi}{d z} \cdot \frac{d z}{d L}.
\end{equation}
First, we differentiate the Z-score term with respect to $L$:
\begin{equation}
    \frac{d z}{d L} = \frac{1}{\sigma_{\widehat{\text{LID}}}}.
\end{equation}
Next, we differentiate the logistic component $\Phi$ with respect to $z$:
\begin{align}
    \frac{d \Phi}{d z} &= \frac{d}{d z} \left( \alpha_{\min} + C (1 + e^{-z})^{-1} \right) \\
    &= C \cdot (-1) \cdot (1 + e^{-z})^{-2} \cdot \frac{d}{d z}(1 + e^{-z}) \\
    &= C \cdot \frac{e^{-z}}{(1 + e^{-z})^2}.
\end{align}
Combining these terms yields the full derivative:
\begin{equation}
    \frac{d \Phi}{d L} = \frac{C}{\sigma_{\widehat{\text{LID}}}} \cdot \frac{e^{-z}}{(1 + e^{-z})^2}.
\end{equation}
We analyze the sign of each component:
\begin{itemize}
    \item The operational range constant $C > 0$.
    \item The standard deviation $\sigma_{\widehat{\text{LID}}} > 0$, assuming the dataset exhibits non-zero geometric variance.
    \item The exponential function $e^{-z} > 0$ for all $z \in \mathbb{R}$.
    \item The denominator $(1 + e^{-z})^2 > 0$.
\end{itemize}
Therefore, the derivative $\frac{d \Phi}{d L} = \frac{C}{\sigma_{\widehat{\text{LID}}}} \frac{e^{-z}}{(1 + e^{-z})^2}$ is strictly positive.
This confirms that the pruning parameter $\alpha$ strictly increases with the local geometric complexity, thereby enforcing a more conservative pruning that retains more navigational edges in high-LID regions.
\end{proof}

\begin{proposition}[Boundedness]\label{proof:bound}
The pruning parameter $\alpha(u)$ derived from the mapping function is strictly bounded within the prescribed operational interval. For any node $u$ with a finite LID estimate:
\begin{equation}
    \alpha_{\min} < \alpha(u) < \alpha_{\max}.
\end{equation}
\end{proposition}

\begin{proof}
Let $S(u)$ denote the logistic component of the mapping function:
\begin{equation}
    S(u) = \frac{1}{1 + \exp(-z(u))}.
\end{equation}
For any finite input $\widehat{\text{LID}}(u)$, the Z-score $z(u)$ is finite. The exponential function maps the real line to the positive real line, i.e., $\exp(-z(u)) \in (0, \infty)$.
Consequently, the denominator lies in the interval $(1, \infty)$. Taking the reciprocal yields the bounds for the logistic component:
\begin{equation}
    0 < S(u) < 1.
\end{equation}
Substituting $S(u)$ back into the definition of $\Phi$:
\begin{equation}
    \alpha(u) = \alpha_{\min} + (\alpha_{\max} - \alpha_{\min}) \cdot S(u).
\end{equation}
Since $(\alpha_{\max} - \alpha_{\min}) > 0$, we can apply the inequality boundaries:
\begin{align}
    \alpha(u) &> \alpha_{\min} + (\alpha_{\max} - \alpha_{\min}) \cdot 0 = \alpha_{\min}, \\
    \alpha(u) &< \alpha_{\min} + (\alpha_{\max} - \alpha_{\min}) \cdot 1 = \alpha_{\max}.
\end{align}
This proves that the topology is strictly confined. 
The pruning behavior never exceeds the relaxation upper limit ($\alpha_{\max}$) and never becomes stricter than the lower limit ($\alpha_{\min}$).
\end{proof}

\begin{lemma}[Local Complexity Lower Bound]
\label{proof:complexity}
Consistent with the complexity bounds established for growth-restricted metrics~\cite{karger2002finding}, consider a query $q$ on a manifold $\mathcal{M}$ of local intrinsic dimensionality $d = \text{LID}(q)$ under the assumptions that the neighbor directions at a node are independent and uniform on the unit sphere, that the step length is proportional to the distance to $q$, and that each greedy step must reduce this distance by a fixed factor $1-\epsilon$ with $\epsilon$ independent of $d$. Then the expected number of distance evaluations $N_{dist}$ per improving step scales exponentially with $d$:
\begin{equation}
    \mathbb{E}[N_{dist}] \ge \Omega\left( \sqrt{d} \cdot \exp(\lambda \cdot d) \right),
\end{equation}
where $\lambda > 0$ depends on $\epsilon$ and the ratio of the step length to the distance.
\end{lemma}

\begin{proof}
Let the local dataset around a query $q$ be modeled as a uniform distribution on a manifold of intrinsic dimension $d$. Consider a greedy routing step where the algorithm is currently at a node $u$ with distance $r = \|u - q\|$.

\paragraph{1. Geometric Condition for Improvement.}
To make progress, the algorithm must identify a neighbor $v$ such that the distance to the query is reduced by a factor $\epsilon$ (where $0 < \epsilon < 1$), i.e., $\|v - q\| \le (1-\epsilon)r$.
Let $s = \|u - v\|$ be the distance between the current node and the candidate neighbor (the step size). Applying the Law of Cosines to triangle $\triangle u v q$, we have:
\begin{equation}
    \|v - q\|^2 = r^2 + s^2 - 2rs \cos \theta,
\end{equation}
where $\theta$ is the angle between the vector $\vec{u q}$ and $\vec{u v}$. The improvement condition $\|v - q\|^2 \le (1-\epsilon)^2 r^2$ implies:
\begin{align}
    r^2 + s^2 - 2rs \cos \theta &\le (1-\epsilon)^2 r^2, \nonumber \\
    2rs \cos \theta &\ge r^2 + s^2 - r^2(1-\epsilon)^2, \nonumber \\
    \cos \theta &\ge \frac{r^2 + s^2 - r^2(1-\epsilon)^2}{2rs}.
\end{align}
For a step size proportional to the distance, i.e., $s = \beta r$ with a constant $\beta > 0$, the right-hand side equals $\frac{1 - (1-\epsilon)^2 + \beta^2}{2\beta}$, a constant that does not depend on $d$. Whenever this constant is below 1, it defines a critical angle $\theta_{max}$ such that any successful neighbor must fall within the cone defined by $0 \le \theta \le \theta_{max}$.

\paragraph{2. Volume Concentration of the Improving Region.}
In the tangent space at $u$, the directions of neighbors are uniformly distributed on the unit hypersphere $\mathbb{S}^{d-1}$. The probability $P_{success}$ of finding a neighbor in the improving cone corresponds to the ratio of the surface area of the spherical cap $\mathcal{C}(\theta_{max})$ to the total surface area of $\mathbb{S}^{d-1}$.
Both areas follow from integrating over the polar angle $\phi$, measured from the direction of $q$. The points of $\mathbb{S}^{d-1}$ at polar angle $\phi$ form a sphere of dimension $d-2$ and radius $\sin \phi$, whose area equals $\sin^{d-2}\phi$ times the area $A_{d-2}$ of the unit sphere $\mathbb{S}^{d-2}$; consequently, the area element at angle $\phi$ is $A_{d-2} \sin^{d-2}\phi \, d\phi$. Integrating this element over $[0, \theta_{max}]$ for the cap and over $[0, \pi]$ for the whole sphere, the constant $A_{d-2}$ cancels in the ratio:
\begin{equation}
    P_{success} = \frac{\text{Area}(\mathcal{C}(\theta_{max}))}{\text{Area}(\mathbb{S}^{d-1})} = \frac{\int_0^{\theta_{max}} \sin^{d-2} \phi \, d\phi}{\int_0^{\pi} \sin^{d-2} \phi \, d\phi}.
    \label{eq:cap_ratio}
\end{equation}
\paragraph{3. Upper Bound on the Success Probability.}
We show that $P_{success} \le \kappa_0\, d^{-1/2} \gamma^{d}$ for every $d \ge 3$, where $\gamma = \sin\theta_{max} < 1$ and the constant $\kappa_0$ does not depend on $d$; this upper bound suffices for the lemma, because $\mathbb{E}[N_{dist}] = 1/P_{success}$. Since $P_{success}$ is the ratio of the two integrals in Eq.~\eqref{eq:cap_ratio}, we bound the numerator from above in Step 3a and the denominator from below in Step 3b before dividing the two bounds in Step 3c.

\emph{Step 3a: numerator.} The numerator of Eq.~\eqref{eq:cap_ratio} is the cap integral $J = \int_0^{\theta_{max}} \sin^{d-2}\phi \, d\phi$, which measures the improving directions; this step shows that $J \le \frac{\sin^{d-1}\theta_{max}}{(d-1)\cos\theta_{max}}$. Because the integrand $\sin^{d-2}\phi$ has no elementary antiderivative, we bound the integral by integration by parts, which needs a factor of the integrand whose antiderivative is known. We construct such a factor as follows. Since $\theta_{max} < \pi/2$, the function $\cos\phi$ is positive on $[0, \theta_{max}]$, which allows us to multiply and divide the integrand by $(d-1)\cos\phi$:
\begin{equation}
    \sin^{d-2}\phi = \frac{1}{(d-1)\cos\phi} \cdot (d-1)\sin^{d-2}\phi \cos\phi.
\end{equation}
The second factor is the derivative of $\sin^{d-1}\phi$, because the chain rule gives $\frac{d}{d\phi}(\sin^{d-1} \phi) = (d-1)\sin^{d-2} \phi \cdot \cos \phi$. We choose the exponent $d-1$ because differentiation lowers it by one to the exponent $d-2$ of the integrand; moreover, the factor $d-1$ that the chain rule produces is the one we multiplied in. The integral therefore becomes
\begin{equation}
    \int_0^{\theta_{max}} \sin^{d-2} \phi \, d\phi = \int_0^{\theta_{max}} \frac{1}{(d-1)\cos \phi} \frac{d}{d\phi}(\sin^{d-1} \phi) \, d\phi.
\end{equation}
Integration by parts states that
\begin{equation}
    \int_a^b f(\phi)\, g'(\phi)\, d\phi = \Bigl[ f(\phi)\, g(\phi) \Bigr]_a^b - \int_a^b f'(\phi)\, g(\phi)\, d\phi,
\end{equation}
and $J$ has exactly the form of the left-hand side with $f(\phi) = \frac{1}{(d-1)\cos\phi}$ and $g(\phi) = \sin^{d-1}\phi$ on $[a, b] = [0, \theta_{max}]$. The right-hand side needs three quantities:
\begin{itemize}
    \item the product $f(\phi)\, g(\phi) = \frac{\sin^{d-1}\phi}{(d-1)\cos\phi}$, which equals $0$ at $\phi = 0$ because $\sin 0 = 0$, and equals $\frac{\sin^{d-1}\theta_{max}}{(d-1)\cos\theta_{max}}$ at $\phi = \theta_{max}$;
    \item the derivative $f'(\phi) = \frac{\sin\phi}{(d-1)\cos^2\phi}$, because the chain rule applied to $(\cos\phi)^{-1}$ gives $\frac{d}{d\phi}(\cos\phi)^{-1} = -(\cos\phi)^{-2} \cdot \frac{d}{d\phi}\cos\phi = -(\cos\phi)^{-2} \cdot (-\sin\phi) = \frac{\sin\phi}{\cos^2\phi}$;
    \item the product $f'(\phi)\, g(\phi) = \frac{\sin\phi \cdot \sin^{d-1}\phi}{(d-1)\cos^2\phi} = \frac{\sin^{d}\phi}{(d-1)\cos^2\phi}$, which we rewrite with $\sin^{d}\phi = \sin^2\phi \cdot \sin^{d-2}\phi$ and $\tan\phi = \sin\phi/\cos\phi$ as $\frac{1}{d-1} \cdot \frac{\sin^2\phi}{\cos^2\phi} \cdot \sin^{d-2}\phi = \frac{1}{d-1} \tan^2\phi \, \sin^{d-2}\phi$.
\end{itemize}
Substituting the three quantities into the right-hand side gives
\begin{equation}
    J = \left[ \frac{\sin^{d-1} \phi}{(d-1)\cos \phi} \right]_0^{\theta_{max}} - \frac{1}{d-1}\int_0^{\theta_{max}} \tan^2\phi \, \sin^{d-2}\phi \, d\phi.
\end{equation}
The last integral is nonnegative, because $\tan^2\phi \ge 0$ and $\sin^{d-2}\phi \ge 0$ on $[0, \theta_{max}]$. Since it enters with a minus sign, dropping it can only increase the right-hand side, which yields the upper bound
\begin{equation}
    J \le \frac{\sin^{d-1} \theta_{max}}{(d-1) \cos \theta_{max}}.
\end{equation}

\emph{Step 3b: denominator.} This step shows that the denominator of Eq.~\eqref{eq:cap_ratio} is at least $1/\sqrt{d}$. The idea is that, because $\sin\phi$ equals $1$ at the equator $\phi = \pi/2$, the integrand $\sin^{d-2}\phi$ stays close to $1$ on a narrow band around the equator. We show that the integrand is at least $\frac{1}{2}$ on the band of half-width $1/\sqrt{d}$; the integral is then at least the width $2/\sqrt{d}$ of the band times $\frac{1}{2}$. The argument has four parts.

(i) Restriction to the band. Because the integrand $\sin^{d-2}\phi$ is nonnegative on $[0, \pi]$, the integral over $[0, \pi]$ is at least the integral over the band $[\pi/2 - 1/\sqrt{d},\ \pi/2 + 1/\sqrt{d}]$, which lies inside $[0, \pi]$ because $1/\sqrt{d} \le 1 < \pi/2$.

(ii) Centering the band at zero. The substitution $\phi = \pi/2 + x$ maps the band to $x \in [-1/\sqrt{d}, 1/\sqrt{d}]$, with $d\phi = dx$ and $\sin(\pi/2 + x) = \sin(\pi/2)\cos x + \cos(\pi/2)\sin x = \cos x$. Parts (i) and (ii) give
\begin{equation}
    \int_0^{\pi} \sin^{d-2} \phi \, d\phi \ge \int_{-1/\sqrt{d}}^{1/\sqrt{d}} \cos^{d-2} x \, dx.
\end{equation}

(iii) Lower bound $\frac{1}{2}$ on the band. For $|x| \le 1/\sqrt{d}$, we show
\begin{equation}
    \cos^{d-2} x \ge \left(1 - \frac{x^2}{2}\right)^{d-2} \ge 1 - \frac{(d-2)\, x^2}{2} \ge 1 - \frac{d-2}{2d} \ge \frac{1}{2}.
\end{equation}
The first inequality follows from $\cos x \ge 1 - x^2/2$, which holds because $1 - \cos x = 2\sin^2(x/2) \le 2\,(x/2)^2 = x^2/2$ by $|\sin y| \le |y|$; since both sides are positive for $|x| \le 1$, raising them to the power $d-2$ keeps the inequality. The second inequality is Bernoulli's inequality $(1 - y)^n \ge 1 - n y$, which holds for every $y \in [0, 1]$ and every real $n \ge 1$; here $y = x^2/2$ lies in $[0, 1]$ because $x^2 \le 1/d$, and $n = d - 2 \ge 1$ because $d \ge 3$, without $d$ being an integer. The third and the fourth inequality use $x^2 \le 1/d$ and $\frac{d-2}{2d} \le \frac{1}{2}$, respectively.

(iv) Integration over the band. Integrating the lower bound $\frac{1}{2}$ from part (iii) over the band of width $2/\sqrt{d}$ gives
\begin{equation}
    \int_0^{\pi} \sin^{d-2} \phi \, d\phi \ge \int_{-1/\sqrt{d}}^{1/\sqrt{d}} \frac{1}{2} \, dx = \frac{2}{\sqrt{d}} \cdot \frac{1}{2} = \frac{1}{\sqrt{d}}.
\end{equation}

\emph{Step 3c: ratio.} A fraction with a positive numerator grows when the numerator grows and when the positive denominator shrinks. Replacing the numerator of Eq.~\eqref{eq:cap_ratio} by its upper bound from Step 3a and the denominator by its lower bound $1/\sqrt{d}$ from Step 3b therefore gives
\begin{equation}
    P_{success} \le \frac{\sin^{d-1} \theta_{max}}{(d-1) \cos \theta_{max}} \cdot \sqrt{d} = \frac{1}{\cos \theta_{max}} \cdot \frac{\sqrt{d}}{d-1} \cdot (\sin \theta_{max})^{d-1}.
\end{equation}
To bring this bound into the form $\kappa_0\, d^{-1/2} \gamma^{d}$, we use $\frac{\sqrt{d}}{d-1} \le \frac{2}{\sqrt{d}}$, which holds for $d \ge 2$ because multiplying both sides by the positive number $\sqrt{d}\,(d-1)$ turns it into $d \le 2(d-1)$. We also write $(\sin \theta_{max})^{d-1} = \gamma^{d}/\gamma$ with $\gamma = \sin \theta_{max}$, which satisfies $0 < \gamma < 1$ because $0 < \theta_{max} < \pi/2$. Finally, we set $\kappa_0 = 2/(\gamma \cos \theta_{max})$ for the factors that do not depend on $d$:
\begin{equation}
    P_{success} \le \frac{1}{\cos\theta_{max}} \cdot \frac{2}{\sqrt{d}} \cdot \frac{\gamma^{d}}{\gamma} = \kappa_0\, d^{-1/2}\, \gamma^{d} = \kappa_0\, d^{-1/2} \exp(d \ln \gamma),
\end{equation}
where the last step writes $\gamma^d = \exp(d \ln \gamma)$. Defining the decay rate $\lambda = -\ln \gamma$, which is positive because $0 < \gamma < 1$ implies $\ln \gamma < 0$, we obtain:
\begin{equation}
    P_{success} \le \kappa_0\, d^{-1/2} \exp(-\lambda \cdot d).
\end{equation}

\paragraph{4. Complexity Bound.}
Under the model of the lemma, each distance evaluation independently finds an improving neighbor with probability $P_{success}$. The number of evaluations until the first success therefore follows a geometric distribution with mean $\mathbb{E}[N_{dist}] = 1/P_{success}$. Step 3 therefore gives $\mathbb{E}[N_{dist}] \ge \kappa_0^{-1} \sqrt{d}\, \exp(\lambda \cdot d)$ for every $d \ge 3$, i.e.,
\begin{equation}
    \mathbb{E}[N_{dist}] \ge \Omega\left( \sqrt{d} \cdot \exp(\lambda \cdot d) \right).
\end{equation}
This confirms that the routing cost is dominated by the exponential term $\exp(\lambda \cdot \text{LID}(q))$, necessitating the adaptive beam width design in MCGI.
\end{proof}

\begin{proposition}[Optimal Budget Allocation]
\label{proof:oba}
To maintain a uniform bound on the routing failure probability $\delta$ across the manifold (i.e., $\mathbb{P}(\text{fail}\;|\;q) \le \delta$ for all $q \in \mathcal{M}$), the search beam width $L(q)$ must grow exponentially with the local intrinsic dimensionality:
\begin{equation}
    L(q) = \Omega\left( \exp\left( \lambda \cdot \text{LID}(q) \right) \right).
\end{equation}
\end{proposition}

\begin{proof}
Consider a beam search with width $L$, and model each of its $L$ expansions as an independent trial that finds a descent step with probability $P_{success}$, as in Definition~\ref{def:iso_recall}. The search fails if all $L$ trials fail, which happens with probability $(1 - P_{success})^L$. A uniform bound $\delta$ on the failure probability therefore requires
\begin{equation}
    (1 - P_{success})^L \le \delta.
\end{equation}
We solve this condition for $L$ in three steps.

(i) Because both sides are positive and the natural logarithm is increasing, the condition is equivalent to
\begin{equation}
    L \cdot \ln(1 - P_{success}) \le \ln \delta.
\end{equation}

(ii) Since $0 < P_{success} < 1$, the factor $\ln(1 - P_{success})$ is negative, and dividing both sides by it reverses the inequality:
\begin{equation}
    L \ge \frac{\ln \delta}{\ln(1 - P_{success})} = \frac{\ln(1/\delta)}{-\ln(1 - P_{success})}.
\end{equation}

(iii) To compare this threshold with $1/P_{success}$, we bound $-\ln(1 - P_{success})$ from above. For $0 \le x < 1$, the Taylor series of the logarithm gives
\begin{equation}
    -\ln(1 - x) = \sum_{k=1}^{\infty} \frac{x^k}{k} \le \sum_{k=1}^{\infty} x^k = \frac{x}{1 - x},
\end{equation}
hence $-\ln(1 - P_{success}) \le P_{success}/(1 - P_{success})$. Inserting this upper bound into the denominator of step (ii) can only lower the threshold; hence every $L$ that meets the bound $\delta$ satisfies
\begin{equation}
    L \ge (1 - P_{success}) \cdot \frac{\ln(1/\delta)}{P_{success}}.
\end{equation}

It remains to bound $P_{success}$ from above. One expansion evaluates at most $R$ neighbors. Step 3 of Proof~\ref{proof:complexity} bounds the probability that one neighbor yields a descent step by $\kappa_0\, d^{-1/2} \exp(-\lambda \cdot d)$, where $d = \text{LID}(q) \ge 3$. The union bound over the at most $R$ neighbors therefore gives
\begin{equation}
    P_{success} \le R\, \kappa_0\, d^{-1/2} \exp(-\lambda \cdot d).
\end{equation}
Since $(1 - P_{success})/P_{success} = 1/P_{success} - 1$, the result of step (iii) reads $L \ge \ln(1/\delta)\,(1/P_{success} - 1)$. The upper bound on $P_{success}$ gives the lower bound $1/P_{success} \ge \sqrt{d}\, \exp(\lambda \cdot d)/(R\, \kappa_0)$. Substituting this lower bound yields, for every $d \ge 3$,
\begin{equation}
    L \ge \ln(1/\delta) \left( \frac{\sqrt{d}\, \exp(\lambda \cdot d)}{R\, \kappa_0} - 1 \right).
\end{equation}
The first term inside the parentheses grows exponentially with $d$, whereas the second term is the constant $1$; consequently, $L(q) = \Omega(\exp(\lambda \cdot \text{LID}(q)))$, which is the claim.
\end{proof}

\begin{proposition}[Connectivity Preservation]
\label{proof:connectivity}
Let $G_{EMST}$ and $G_{RNG}$ denote the Euclidean Minimum Spanning Tree and the Relative Neighborhood Graph, respectively. For any configuration of points with distinct pairwise distances, provided that $\alpha(u) \ge 1.0$ for all $u \in V$, that the pruning test of each node runs over all other points, and that no degree bound truncates the neighbor lists, the graph $G_{MCGI}$ satisfies the inclusion hierarchy:
\begin{equation}
    E_{EMST} \subseteq E_{RNG} \subseteq E_{MCGI}.
\end{equation}
Consequently, $G_{MCGI}$ is connected.
\end{proposition}

\begin{proof}
The proof proceeds in two steps. First, we invoke the classical result established by Toussaint~\cite{TOUSSAINT1980261}, which proves that the Relative Neighborhood Graph (RNG) is a supergraph of the Euclidean Minimum Spanning Tree (EMST) for any finite set of points in a metric space:
\begin{equation}
    E_{EMST} \subseteq E_{RNG}.
\end{equation}
Since the EMST connects all vertices in a single component, $G_{RNG}$ is necessarily connected.

Second, we show that $E_{RNG} \subseteq E_{MCGI}$. In the RNG, an edge between $u$ and $v$ is pruned if and only if a witness node falls within the intersection of two spheres centered at $u$ and $v$ with radius $d(u, v)$. We denote this geometric intersection as the exclusion region $\mathcal{R}_{RNG}$. Therefore, an edge exists in the RNG if and only if $\mathcal{R}_{RNG}$ contains no data points.
In \ours{}, the candidates of $u$ are processed in increasing order of their distance to $u$, with a previously selected neighbor as the witness; hence, the witness satisfies $d(u, n) < d(u, v)$ and $d(n, v) \le \frac{1}{\alpha(u)} d(u, v)$. Because the algorithm enforces $\alpha(u) \ge 1.0$, the second distance is at most $d(u, v)$. Consequently, the exclusion region of \ours{} is a geometric subset of the RNG exclusion region:
\begin{equation}
    \mathcal{R}_{MCGI} \subseteq \mathcal{R}_{RNG}.
\end{equation}
This subset relationship of exclusion regions creates an inverse superset relationship for the preserved edges. If an edge survives the RNG condition, its $\mathcal{R}_{RNG}$ is empty. Since $\mathcal{R}_{MCGI}$ is contained entirely within $\mathcal{R}_{RNG}$, it must also be empty. Thus, any edge present in the RNG is guaranteed to be preserved in \ours{}, yielding the inclusion hierarchy of edge sets:
\begin{equation}
    E_{RNG} \subseteq E_{MCGI}.
\end{equation}
By combining this with the classical result that $E_{EMST} \subseteq E_{RNG}$ derived in \cite{TOUSSAINT1980261}, we establish the complete global connectivity chain.
\end{proof}

\begin{proposition}[Suboptimality of Fixed Beam]
\label{proof:fixed_beam}
For any fixed beam width $\bar{L}$, the probability that the search succeeds at a query $q$ is at most $\ln(1/\delta)\, \bar{L}/L^*(q)$, where $L^*(q)$ is the iso-recall list size of Definition~\ref{def:iso_recall}. Consequently, this probability decays exponentially with $\mathrm{LID}(q)$.
\end{proposition}

\begin{proof}
Consider a beam search with fixed width $\bar{L}$. Under the model of Definition~\ref{def:iso_recall}, each of its $\bar{L}$ expansions succeeds independently with probability $P_{\mathrm{success}}(q)$. The failure probability at query $q$ is therefore
\begin{equation}
    P(\mathrm{fail} \mid q) = \left(1 - P_{\mathrm{success}}(q)\right)^{\bar{L}}.
\end{equation}
Bernoulli's inequality $(1-x)^{n} \ge 1 - n x$ for $x \in [0, 1]$ and integer $n \ge 1$ gives the first claim:
\begin{equation}
    1 - P(\mathrm{fail} \mid q) \le \bar{L}\, P_{\mathrm{success}}(q).
\end{equation}
Definition~\ref{def:iso_recall} sets $L^*(q) = \ln(1/\delta)/P_{\mathrm{success}}(q)$. Substituting the resulting value $P_{\mathrm{success}}(q) = \ln(1/\delta)/L^*(q)$ into the bound above gives the claim:
\begin{equation}
    1 - P(\mathrm{fail} \mid q) \le \ln(1/\delta)\, \frac{\bar{L}}{L^*(q)}.
\end{equation}
For the exponential decay, Eq.~\eqref{eq:oracle_schedule} gives $L^*(q) \ge C_\delta \exp(\lambda \cdot \mathrm{LID}(q))$ with $C_\delta = \ln(1/\delta)/(R\, \kappa_0)$. Inserting this lower bound into the denominator cancels the factor $\ln(1/\delta)$:
\begin{equation}
    1 - P(\mathrm{fail} \mid q) \le \ln(1/\delta)\, \frac{\bar{L}}{C_\delta \exp(\lambda \cdot \mathrm{LID}(q))} = \bar{L}\, R\, \kappa_0 \exp\!\left(-\lambda \cdot \mathrm{LID}(q)\right),
    \label{eq:recall_gap}
\end{equation}
which decays exponentially with $\mathrm{LID}(q)$.
\end{proof}

\subsection{Additional Experimental Details}
\label{appendix:setup}

\paragraph{Experiment Platform}
All experiments run on a cloud virtual machine with 192 virtual cores of Intel Xeon Platinum 8468 and 960 GiB of RAM. The operating system is Ubuntu 24.04 LTS, with all code compiled by GCC 13.3 at optimization level -O3. The disk-resident indexes reside on a 17.6 TB virtual block device (virtio) of the virtual machine, which the DiskANN search engine reads through Linux asynchronous I/O with direct I/O, so that the operating system page cache does not serve graph nodes. Every search uses 32 threads and a beam width of 2, with the page cache dropped before each search run. All indexes are built on the CPU cores with the DiskANN builder; the SIFT1B and Deep1B builds use a memory budget of 300 GiB and 192 threads.

\paragraph{Datasets}
We evaluate our method on three standard million-scale benchmarks whose dimensionality ranges from 100 to 960 and on two billion-scale benchmarks, so that the datasets cover both high ambient dimensionality and large scale. In the following descriptions, $N$ denotes the number of base vectors and $D$ represents the vector dimensionality:
\begin{itemize}
    \item SIFT1M~\cite{jegou2011product} ($N=10^6, D=128$): Local image descriptors with 10{,}000 queries, searched with Euclidean distance ($L_2$).
    \item GloVe-100~\cite{pennington2014glove} ($N=1.18 \times 10^6, D=100$): Word embedding vectors with 10{,}000 queries. Following standard practice, we normalize the vectors to unit length and use Euclidean distance as a proxy for cosine similarity.
    \item GIST1M~\cite{jegou2011product} ($N=10^6, D=960$): Global image descriptors with 1{,}000 queries, which have the highest ambient dimensionality and the widest spread of LID values among the million-scale datasets.
    \item SIFT1B~\cite{jegou2011searching} ($N=10^9, D=128$): One billion local image descriptors stored as 8-bit unsigned integers, with 10{,}000 queries and the official ground truth over the full billion vectors.
    \item Deep1B~\cite{babenko2016efficient} ($N=10^9, D=96$): One billion image descriptors from a convolutional neural network, reduced to 96 dimensions by PCA, normalized to unit length, and stored as 32-bit floats, with 10{,}000 queries. The official ground truth over the full billion vectors is wrong for three queries (indices 449, 2382, and 2809)\footnote{\url{https://github.com/harsha-simhadri/big-ann-benchmarks/issues/353}}, which we exclude.
\end{itemize}
The calibration queries of the query-adaptive search list are disjoint from the evaluation queries. For SIFT1M and GIST1M, they are the first 10{,}000 vectors of the learning set of each dataset. GloVe-100 has no learning set; its calibration queries are 10{,}000 base vectors drawn with a fixed seed, each excluded from its own ground truth. The ground truth of every calibration query consists of its exact 100 nearest neighbors. For SIFT1B and Deep1B, the calibration queries are the even-indexed half of the official queries, whereas the evaluation uses the odd-indexed half, which leaves 5{,}000 evaluation queries on SIFT1B and 4{,}998 on Deep1B.

\paragraph{Baselines}
We compare \ours{} against DiskANN (Vamana)~\cite{subramanya2019diskann} on every dataset and, on GIST1M, additionally against PipeANN~\cite{osdi25pipeann, fast26odinann}, CSPG~\cite{NEURIPS2024_bab1486c}, HNSW~\cite{malkov2018efficient}, and IVF-Flat~\cite{johnson2019billion}. DiskANN is the standard disk-resident graph index; \ours{} and DiskANN search the same index file with the same compressed vectors, the same beam width, and the same requested list sizes. The comparison with DiskANN therefore isolates the query-adaptive search list of \ours{}, since the index, the I/O path, and the distance computation are identical. Each of the four GIST1M baselines searches the 1{,}000 official queries with 32 threads once per setting, with the following configurations.
\begin{itemize}
    \item PipeANN~\cite{osdi25pipeann, fast26odinann} (pipelined search with \texttt{io\_uring}) searches the DiskANN index files of \ours{} with requested list sizes from 10 to 400 and I/O widths 4 and 32, of which width 4 gives the higher throughput and enters Table~\ref{tab:gist_compare}. Its entry-point index is built from a 1\% sample of the base vectors; in addition, its per-vector PQ limit is raised from 256 to 512 bytes so that it loads the 322-byte codes of the GIST1M indexes.
    \item CSPG~\cite{NEURIPS2024_bab1486c} uses the official implementation with the settings of its authors for GIST1M, namely two partitions of Vamana subgraphs, a construction list of 128, $\alpha = 1.2$, and a duplication ratio of 0.5, with $R = 32$ and $R = 48$. Its parallel search assigns queries to threads in blocks of 64; the 1{,}000 GIST1M queries therefore occupy at most 16 of the 32 threads.
    \item HNSW~\cite{malkov2018efficient} uses hnswlib with $M = 16$, which bounds the degree of the base layer by 32, and a construction list of 200.
    \item IVF-Flat~\cite{johnson2019billion} uses Faiss with 4{,}096 lists and full-precision vectors.
\end{itemize}

\paragraph{Evaluation Metrics}
We adhere to the standard evaluation protocol for ANN search and measure Recall@10 against QPS (queries per second). Each method yields one Recall@10 and QPS point per requested list size. The throughput gain at a target recall is the ratio of the QPS of the two methods, where each QPS value is linearly interpolated along the recall axis of its curve.

\paragraph{Build Configuration}
Table~\ref{tab:build_params} summarizes the dataset specifications and the indexing parameters of the index that \ours{} and DiskANN search on each dataset, where the query count is the number of evaluation queries. $R$ denotes the maximum outdegree of each node in the graph, and $L_{build}$ is the size of the candidate list during index construction. Every index uses the standard pruning parameter $\alpha = 1.2$ of DiskANN. $m_{PQ}$ specifies the number of bytes of the Product Quantization (PQ) code of each vector, which follows from the PQ memory budget of the build.

\begin{table}[t]
\centering
\caption{Dataset specifications and indexing parameters.}
\label{tab:build_params}
\begin{tabular}{lccccccc}
\toprule
Dataset & $N$ & $D$ & Queries & $R$ & $L_{build}$ & $m_{PQ}$ (Byte) & Data Type\\
\midrule
SIFT1M    & $10^6$ & 128 & 10{,}000 & 32 & 50  & 107 & float32 \\
GloVe-100 & $1.18 \times 10^6$ & 100 & 10{,}000 & 64 & 100 & 90  & float32 \\
GIST1M    & $10^6$ & 960 & 1{,}000  & 32 & 100 & 322 & float32 \\
SIFT1B    & $10^9$ & 128 & 5{,}000  & 32 & 50  & 31  & uint8   \\
Deep1B    & $10^9$ & 96  & 4{,}998  & 32 & 50  & 31  & float32 \\
\bottomrule
\end{tabular}
\end{table}

\paragraph{Search Configuration}
Table~\ref{tab:search_params} lists the LID statistics of each dataset and the parameters of the query-adaptive search list. $\mu_{LID}$ and $\sigma_{LID}$ denote the mean and standard deviation of the maximum-likelihood LID estimates of the base vectors with 20 neighbors, which we compute for the million-scale datasets only. The column for the list size $\bar{L}$ gives the range of requested list sizes over which the curves of \ours{} and DiskANN are measured on each dataset. On the million-scale datasets, the range $10 \to 400$ consists of the 14 values 10, 15, 20, 25, 30, 40, 50, 60, 80, 100, 150, 200, 300, and 400; SIFT1B adds the four values 500, 600, 800, and 1{,}000. Deep1B adds these four values together with 1{,}200, 1{,}500, and 2{,}000. The billion-scale datasets need these longer lists because DiskANN reaches high recall on them only with long lists. The checkpoint of Eq.~\eqref{eq:adaptive_list} follows $n_0 = 24$ expansions; the candidate-list estimator uses the $m = 64$ best candidates; the competition count uses $\epsilon = 0.2$. The list size is bounded from below by the result size $k = 10$ and from above by $c\bar{L}$ with $c = 8$. The strength $\gamma$ is chosen from $\{0.5, 0.7, 1.0, 1.3\}$ on the calibration queries of each dataset.

\begin{table}[t]
\centering
\small 
\caption{LID statistics and search parameters.}
\label{tab:search_params}
\begin{tabular}{lccccc}
\toprule
Dataset & $\mu_{LID}$ (Mean) & $\sigma_{LID}$ (Std) & List size $\bar{L}$ & $\gamma$ & Threads \\
\midrule
SIFT1M    & 16.1 & 9.2  & $10 \to 400$ & 0.7 & 32 \\
GloVe-100 & 35.3 & 18.1 & $10 \to 400$ & 1.0 & 32 \\
GIST1M    & 48.9 & 25.9 & $10 \to 400$ & 0.7 & 32 \\
SIFT1B    & -- & -- & $10 \to 1000$ & 0.7 & 32 \\
Deep1B    & -- & -- & $10 \to 2000$ & 0.7 & 32 \\
\bottomrule
\end{tabular}
\end{table}

\begin{figure*}[t]
    \centering
    \includegraphics[width=\textwidth]{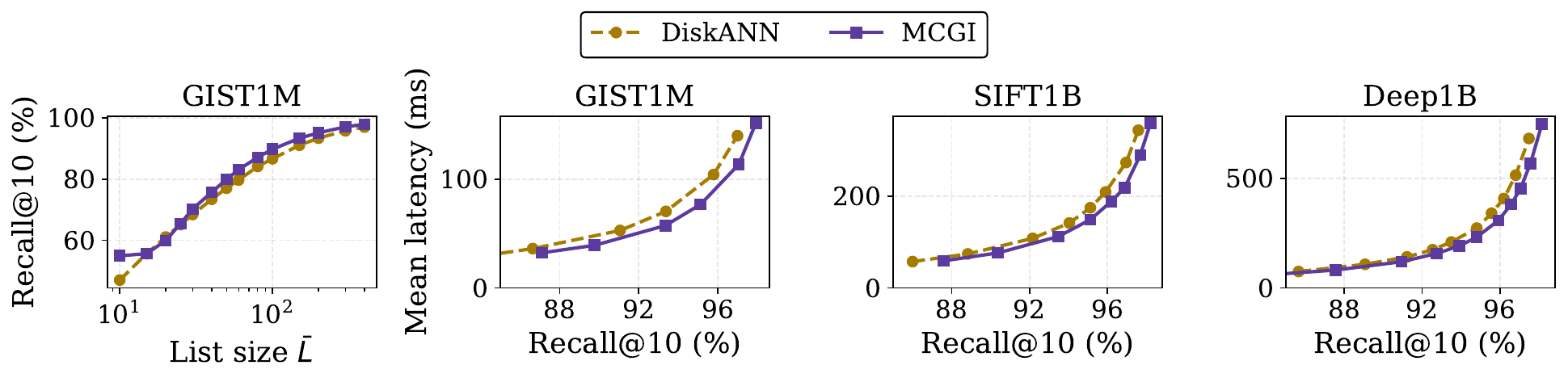}
    \caption{Recall against the requested list size, and mean latency against recall.}
    \label{fig:latency}
\end{figure*}

\paragraph{Latency}
Figure~\ref{fig:latency} plots the recall of \ours{} and DiskANN against the requested list size $\bar{L}$ on GIST1M and the mean query latency against recall on GIST1M, SIFT1B, and Deep1B. The recall of \ours{} closely follows that of DiskANN at every requested list size, so $\bar{L}$ remains the single knob of the search. At matched recall, \ours{} lowers the mean latency, at 95\% recall from 93 to 75~ms on GIST1M and from 290 to 245~ms on Deep1B, and at 90\% recall from 86 to 74~ms on SIFT1B.

\paragraph{Offline Cost}
Table~\ref{tab:offline_cost} lists the offline cost of each index, where every build runs alone on the platform with 192 threads and the calibration runs alone with 32 search threads. For the million-scale datasets, the table reports a separate rebuild with the same parameters, because the evaluated indexes were built while other jobs shared the machine. The training of the PQ codebooks dominates the build of the million-scale indexes, whereas the graph construction over the shards and their merge dominates at one billion vectors. The only offline step that the query-adaptive search list adds is the calibration of Eq.~\eqref{eq:adaptive_list}, which searches the calibration queries once at each list size of a fixed grid and then fits the weights by ridge regression within seconds. The grid has 14 list sizes from 10 to 400 for the million-scale datasets and SIFT1B, and 18 list sizes from 10 to 1{,}000 for Deep1B. Calibration takes about 3 minutes on each million-scale dataset, and 1.6 and 4.8 minutes on SIFT1B and Deep1B, which is at most 1.5\% of the build time of a billion-scale index.

\begin{table}[t]
\centering
\small
\caption{Offline cost in seconds: index build stages and calibration of the query-adaptive search list.}
\label{tab:offline_cost}
\begin{tabular}{lccccc}
\toprule
Dataset & PQ training & Graph & Disk layout & Total build & Calibration \\
\midrule
SIFT1M    & 195   & 6       & 1   & 203       & 174 \\
GloVe-100 & 400   & 15      & 1   & 416       & 168 \\
GIST1M    & 1{,}678 & 82    & 4   & 1{,}764   & 174 \\
SIFT1B    & 2{,}775 & 14{,}117 & 290 & 17{,}225 & 94  \\
Deep1B    & 3{,}373 & 15{,}029 & 836 & 19{,}317 & 289 \\
\bottomrule
\end{tabular}
\end{table}

\subsection{Limitations}
\label{appendix:limitations}

All throughput gains in Section~\ref{sec:experiments} come from the query-adaptive search list on indexes built with the standard Vamana pruning rule. The construction-time rule of Section~\ref{subsec:mapping} relaxes the pruning test of high-LID nodes; however, under the degree bound of a disk-resident index, nodes typically fill their neighbor lists with either rule. The construction-time rule therefore changes mainly the choice of edges of a node rather than their number, which is why the evaluation attributes no gain to it. The connectivity guarantee of Proposition~\ref{prop:connectivity} likewise holds only without the degree bound. Similarly, because the analysis of Section~\ref{sec:theory} models each beam step as an independent trial under a uniform local model of the data, its bounds describe trends rather than exact recall values.

The query-adaptive list also carries two costs that its mean throughput hides. First, because hard queries receive lists up to eight times longer than the requested list, the 99.9th-percentile latency at matched recall rises while the mean latency falls; on GIST1M at 95\% recall, for example, the tail latency grows from 119 to 181~ms. Second, every index needs its own calibration with a few thousand queries from the query distribution and their exact nearest neighbors, since the fitted weights hold only for the index on which they were fitted.

\subsection{Algorithm Pseudo-Code}
\label{appendix:alg}

Algorithm~\ref{alg:mcgi} outlines the offline construction process of the indexing structure. The procedure is divided into two sequential phases. 
\begin{itemize}
    \item The first phase focuses on geometric calibration by calculating the Local Intrinsic Dimensionality for all vectors in the dataset in parallel. After obtaining the global mean and standard deviation of these estimates, the algorithm assigns a personalized pruning parameter $\alpha_u$ to each node. This parameter is derived using a logistic function that maps the standardized geometric complexity of the local neighborhood into a bounded operational range. 
    \item The second phase handles topology refinement. Starting from a random graph, the algorithm iteratively optimizes the connections. For each node, it executes a greedy search to identify a pool of candidate neighbors. It then evaluates these candidates alongside existing neighbors using a dynamic pruning criterion. A candidate is pruned if a previously accepted neighbor falls within a restricted exclusion region defined by the precomputed $\alpha_u$. Survivors are added to the updated neighbor list until the degree limit is reached.
\end{itemize}

\begin{algorithm}[h!]
   \caption{Manifold-Consistent Graph Indexing (MCGI)}
   \label{alg:mcgi}
\begin{algorithmic}
   \STATE Input: Dataset $X$, Max Degree $R$, Beam Width $L$
   \STATE Output: Optimized Graph $G$

   \STATE
   \STATE // Phase 1: Geometric Calibration
   \STATE $\mathcal{L} \leftarrow \text{ParallelEstimateLID}(X)$
   \STATE $\mu \leftarrow \text{Mean}(\mathcal{L})$
   \STATE $\sigma \leftarrow \text{StdDev}(\mathcal{L})$

   \FOR{each node $u \in V$ in parallel}
       \STATE $z_u \leftarrow (\mathcal{L}[u] - \mu) / \sigma$
       \STATE $\alpha_u \leftarrow \alpha_{\min} + (\alpha_{\max} - \alpha_{\min}) / (1 + \exp(-z_u))$
   \ENDFOR

   \STATE 
   \STATE // Phase 2: Topology Refinement
   \STATE $G \leftarrow \text{RandomGraph}(X, R)$
   
   \FOR{$iter \leftarrow 1$ to $MaxIter$}
       \FOR{each node $u \in G$ in parallel}
           \STATE $\mathcal{C} \leftarrow \text{GreedySearch}(u, G, L)$
           \STATE $\mathcal{N}_{new} \leftarrow \emptyset$
           
           \FOR{$v \in \text{SortByDistance}(\mathcal{C} \cup \mathcal{N}(u))$}
               \STATE $pruned \leftarrow \text{False}$
               \FOR{$n \in \mathcal{N}_{new}$}
                   \IF{$\alpha_u \cdot d(n, v) \le d(u, v)$}
                       \STATE $pruned \leftarrow \text{True}$; break
                   \ENDIF
               \ENDFOR
               \IF{not $pruned$ \AND $|\mathcal{N}_{new}| < R$}
                   \STATE $\mathcal{N}_{new}.\text{add}(v)$
               \ENDIF
           \ENDFOR
           \STATE $\mathcal{N}(u) \leftarrow \mathcal{N}_{new}$
       \ENDFOR
   \ENDFOR
\end{algorithmic}
\end{algorithm}

Algorithm~\ref{alg:online_mcgi} presents an optimized online variant designed to handle massive datasets where exhaustive geometric profiling is computationally prohibitive. 
\begin{itemize}
    \item It begins with a fast approximation phase that samples a small subset of the data. By evaluating this subset, the algorithm quickly bootstraps the global mean and standard deviation of the underlying probability distribution. This step provides a coarse but efficient estimate of the overall geometric complexity, enabling the algorithm to initialize the pruning parameters without incurring the cost of a full-dataset pass.
    \item The second phase merges the geometric estimation directly into the iterative topology-refinement loop. Instead of computing all complexities beforehand, the algorithm dynamically estimates the node-specific dimensionality on the fly using the candidate set discovered during the greedy-search step. This real-time estimate is immediately used to calculate the specific pruning parameter $\alpha_u$. The algorithm then applies this dynamic parameter to the pruning criterion, determining which edges to preserve and which to discard. This approach ensures the graph connectivity adapts to the manifold geometry progressively as the quality of the candidate neighbors improves with each iteration.
\end{itemize}

\begin{algorithm}[h!]
   \caption{Online Manifold-Consistent Graph Indexing}
   \label{alg:online_mcgi}
\begin{algorithmic}
   \STATE Input: Dataset $X$, Max Degree $R$, Beam Width $L$, Sample Rate $S$
   \STATE Output: Optimized Graph $G$

   \STATE
   \STATE // Phase 1: Bootstrap Statistics (Fast Approximation)
   \STATE $X_{sample} \leftarrow \text{RandomSample}(X, S)$
   \STATE $\mathcal{L}_{sample} \leftarrow \text{EstimateLID}(X_{sample})$ \COMMENT{Compute LID on subset}
   \STATE $\mu \leftarrow \text{Mean}(\mathcal{L}_{sample})$, $\sigma \leftarrow \text{StdDev}(\mathcal{L}_{sample})$

   \STATE 
   \STATE // Phase 2: Online Topology Refinement
   \STATE $G \leftarrow \text{RandomGraph}(X, R)$
   
   \FOR{$iter \leftarrow 1$ to $MaxIter$}
       \FOR{each node $u \in G$ in parallel}
           \STATE $\mathcal{C} \leftarrow \text{GreedySearch}(u, G, L)$
           
           \STATE // Online LID Estimation \& Parameter Adaptation
           \STATE $\hat{\ell}_u \leftarrow \text{ComputeLID}(u, \mathcal{C})$ \COMMENT{Estimate from neighbors}
           \STATE $z_u \leftarrow (\hat{\ell}_u - \mu) / \sigma$
           \STATE $\alpha_u \leftarrow \alpha_{\min} + (\alpha_{\max} - \alpha_{\min}) / (1 + \exp(-z_u))$

           \STATE $\mathcal{N}_{new} \leftarrow \emptyset$
           
           \FOR{$v \in \text{SortByDistance}(\mathcal{C} \cup \mathcal{N}(u))$}
               \STATE $pruned \leftarrow \text{False}$
               \FOR{$n \in \mathcal{N}_{new}$}
                   \STATE // Pruning with dynamic $\alpha_u$
                   \IF{$\alpha_u \cdot d(n, v) \le d(u, v)$}
                       \STATE $pruned \leftarrow \text{True}$; break
                   \ENDIF
               \ENDFOR
               \IF{not $pruned$ \AND $|\mathcal{N}_{new}| < R$}
                   \STATE $\mathcal{N}_{new}.\text{add}(v)$
               \ENDIF
           \ENDFOR
           \STATE $\mathcal{N}(u) \leftarrow \mathcal{N}_{new}$
       \ENDFOR
   \ENDFOR
\end{algorithmic}
\end{algorithm}

\end{document}